\documentclass[fleqn,usenatbib]{mnras}

\usepackage{newtxtext,newtxmath}
\usepackage[T1]{fontenc}
\usepackage{ae,aecompl}

\usepackage[utf8]{inputenc}
\usepackage{amsmath}
\usepackage{graphicx}
\usepackage{amsfonts}
\usepackage{xcolor}
\usepackage{verbatim} 
\usepackage{caption}
\usepackage{subcaption}
\usepackage{ulem}

\newcommand{\orcid}[1]{\href{https://orcid.org/#1}{\includegraphics[scale=0.08]{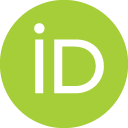}}}

\newcommand\ddfrac[2]{\frac{\displaystyle #1}{\displaystyle #2}}

\newcommand{\mvir}{M_{\rm 200c}}
\newcommand{\mvirhost}{\mvir^{\rm host}}
\newcommand{\rvir}{R_{\rm 200c}}
\newcommand{\rvirhost}{\rvir^{\rm host}}

\newcommand{\mcgas}{M_{\rm ColdGas}}
\newcommand{\mcgassat}{\mcgas^{\rm sat}}
\newcommand{\mcgassub}{\mcgas^{\rm subhalo}}
\newcommand{\mstar}{M_\star}
\newcommand{\mstarsat}{\mstar^{\rm sat}}
\newcommand{\rgal}{R_{\rm gal}}
\newcommand{\dsathost}{d_{\rm sat}^{\rm host}}
\newcommand{\sublink}{\textsc{sublink}}
\newcommand{\sublinkgal}{\sublink\_\textsc{gal} }
\newcommand{\subfind}{\textsc{subfind} }

\newcommand{\msun}{{\rm M}_\odot}

\graphicspath{{figures/}}
\DeclareGraphicsExtensions{.pdf,.png,.svg,.gif}
\DeclareCaptionFormat{cont}{#1 (cont.)#2#3\par}

\title[When, where, and for how long RPS occurs]{Jellyfish galaxies with the IllustrisTNG simulations --  When, where, and for how long does ram pressure stripping of cold gas occur?}
\author[Rohr et al.]{Eric Rohr$^{1}$\thanks{Contact e-mail: \href{mailto:rohr@mpia.de}{rohr@mpia.de}}\orcid{0000-0002-9183-5593},
Annalisa Pillepich$^{1}$\orcid{0000-0003-1065-9274}, Dylan Nelson$^2$\orcid{0000-0001-8421-5890}, Elad Zinger$^{3,1}$\orcid{0000-0002-6316-3996}, Gandhali D. Joshi$^4$\orcid{0000-0003-4665-0765},
\newauthor 
and Mohammadreza Ayromlou$^2$\orcid{0000-0003-3783-2321}
\\
\\
$^{1}$Max-Planck-Institut f{\"u}r Astronomie, K{\"o}nigstuhl 17, D-69117 Heidelberg, Germany\\
$^{2}$Zentrum f{\"u}r Astronomie der Universit{\"a}t Heidelberg, ITA, Albert Ueberle Str. 2, D-69120 Heidelberg, Germany\\
$^{3}$Centre for Astrophysics and Planetary Science, Racah Institute of Physics, The Hebrew University, Jerusalem 91904, Israel \\
$^{4}$Department of Physics and Astronomy, University College London, Gower St, London WC1E 6BT, UK\\
}

\date{}

\pubyear{2023}

\begin{document}
\label{firstpage}
\pagerange{\pageref{firstpage}--\pageref{lastpage}}
\maketitle

\begin{abstract}

Jellyfish galaxies are prototypical examples of satellite galaxies undergoing strong ram pressure stripping (RPS). We analyze the evolution of 512 unique, first-infalling jellyfish galaxies from the TNG50 cosmological simulation. These have been visually inspected to be undergoing RPS sometime in the past 5 billion years~(since~$z=0.5$), have satellite stellar masses~$\mstarsat\sim10^{8-10.5}\,\msun$, and live in hosts with~$\mvir\sim10^{12-14.3}\,\msun$~at~$z=0$. We quantify the cold gas~($T\leq10^{4.5}$~K)~removal using the tracer particles, confirming that for these jellyfish, RPS is the dominant driver of cold gas loss after infall. Half of these jellyfish are completely gas-less by~$z=0$, and these galaxies have earlier infall times and smaller satellite-to-host mass ratios than their gaseous counterparts. RPS can act on jellyfish galaxies over long time scales of~$\approx1.5-8$~Gyr. Jellyfish in more massive hosts are impacted by RPS for a shorter time span and, at a fixed host mass, jellyfish with less cold gas at infall and lower stellar masses at~$z=0$~have shorter RPS time spans. While RPS may act for long periods of time, the peak RPS period -- where at least 50~per~cent of the total RPS occurs -- begins within~$\approx1$~Gyr of infall and lasts~$\lesssim2$~Gyr. During this period, the jellyfish are at host-centric distances~$\sim0.2-2\rvir$, illustrating that much of RPS occurs at large distances from the host galaxy. Interestingly, jellyfish continue forming stars until they have lost $\approx98$~per~cent of their cold gas. For groups and clusters in TNG50~$(\mvirhost\sim10^{13-14.3}\,\msun)$, jellyfish galaxies deposit more cold gas~($\sim10^{11-12}\,\msun$)~into halos than exist in them at~$z=0$, demonstrating that jellyfish, and in general satellite galaxies, are a significant source of cold gas accretion.

\end{abstract}


\section{Introduction} \label{sec:intro}

At a fixed galaxy stellar mass, observations show that there are a number of differences between field and satellite galaxies (satellites for short). Namely with the Sloan Digital Sky Survey (SDSS), it has been shown that the population of satellites has a higher quenched fraction, lower (specific) star-formation rates (SFR, or sSFR), and redder colors compared to central galaxies of the same stellar mass \citep{Peng2010,Peng2012,Wetzel2012}. Moreover, satellite galaxies exhibit on average lower neutral HI gas fractions, elevated gas metallicites, reduced circumgalactic X-ray emission, and suppressed active galactic nucleus (AGN) activity compared to their mass-matched analogs in the field \citep{Giovanelli1985,Brown2016,Maier2019,Maier2019b}. 

These observational trends suggest that, in addition to the secular processes of galaxy evolution, satellite galaxies undergo additional environmental phenomena. It is generally accepted that ram-pressure stripping (RPS) is one of the most impactful among such environmental phenomena \citep[][see \citealp{Boselli2022} for a recent review]{Gunn1972}. 

Ram pressure is proportional to $\rho v^2$, where $\rho$ is the density of the surrounding ambient medium, and $v$ is the relative velocity of the infalling galaxy (or a given parcel of gas) and the ambient medium. This effect is expected to increase with host mass ($\mvir$) because satellites in more massive hosts tend to fall in with higher velocities and more massive hosts tend to have denser circumgalactic media (CGM)\footnote{Throughout this paper, we define the CGM to be the entire multiphase gaseous medium around central galaxies regardless of their stellar or total host mass, unless explicitly referred to as intragroup medium (IGrM) for galaxy groups ($\mvir \sim 10^{13-14}\, \msun$) or intracluster medium (ICM) for clusters ($\mvir \sim 10^{14-14.3}\, \msun$).}, also depending on the stellar and AGN feedback of the central galaxy. Moreover for a given host, this pressure should increase with decreasing distance both because the surrounding medium is denser at smaller radii, and galaxies move faster when they are deeper into their hosts' potential wells. These expected results are broadly consistent with observations \citep[e.g.,][]{Maier2019b,Roberts2019} With respect to removing single parcels of gas from the infalling satellite, RPS acts against the satellite's gravitational restoring force, dominated by the stellar body. Consequently the effectiveness of RPS is expected to increase with decreasing satellite stellar mass. 

For a given satellite galaxy, ram pressure first strips the hot or less gravitationally-bound gas, a feature that has been inferred observationally \citep{Balogh2000} and assumed in semi-analytic models \citep{Cole2000,Somerville2008,Lagos2018,Ayromlou2019}. With respect to the satellite's intersteller medium (ISM), RPS is thought to work outside-in, as observationally inferred via truncated disks \citep{Warmels1988c,Cayatte1990,Cayatte1994,Vollmer2001,Lee2022b} and leading to outside-in quenching \citep[][contra: \citealp{Wang2022b}]{Schaefer2017,Schaefer2019,Bluck2020,Vulcani2020,Wang2023}.

Ram pressure is also thought to compress the satellite's gas, especially on the galaxy's leading edge. This is inferred to cause temporary periods of enhanced star formation \citep{Gavazzi2001,Vulcani2018,Roberts2020,Grishin2021,Roberts2022} and AGN activity \citep[][contra: \citealp{Roman-Oliveira2019}]{Poggianti2017b,Maier2022,Peluso2022}. In turn, the feedback from star-formation and AGN may lower the binding energy of the ISM gas, potentially facilitating RPS \citep{Garling2022}. Thus the physical mechanism responsible for the loss of satellite ISM gas is likely a combination of RPS and stellar/AGN-driven outflows. However, despite these temporary periods of enhanced star-formation and AGN activity, RPS ultimately leads to the removal of ISM gas and to the quenching {\it en masse} of satellites \citep[e.g.,][see \citealp{Cortese2021} for a recent review]{Wetzel2013,Maier2019,Boselli2022}. We note, however, that the timescales related to environmental quenching are highly debated, ranging from short $\lesssim 500$~Myr to long $\gtrsim 4$~Gyr times, typically but not always measured from the first $\rvir$ crossing \citep[][and references therein]{Cortese2021}. 

Conversely, satellite galaxies are not only affected by their environment, but they have the potential to perturb the ambient medium in a number of ways. First, the bulk motion of the satellites is thought to affect the CGM kinematics by inducing turbulence and by bringing in gravitational energy, which heats the CGM via dynamical friction and shocks \citep[e.g.,][]{Dekel2008}. As the infalling galaxies may travel faster than the ambient medium's sound speed, some satellites are also expected to create bow shocks in CGM \citep{Yun2019}. This shock and the induced turbulence may act as perturbations, triggering the warm/hot $T\sim10^{6-8}$~K CGM to cool into $T\sim 10^{4-5}$~K clouds. Moreover, the gas that has been ram-pressure stripped, namely the satellite's cold ISM, is expected to be deposited into the host's halo. For groups and clusters with many satellite galaxies, there could be a substantial amount of accreted halo gas originating from the stripped satellites. However, this has never been quantified. Finally, while currently still highly debated, such cold gas clouds in the CGM, regardless of their origin, could be long-lived \citep[e.g.,][]{Li2020,Sparre2020,Gronke2022,Fielding2022}, and satellite-induced cold gas clouds may be a source of cold gas found in the CGM today \citep[][]{Nelson2020,Rodriguez2022}.

Observed satellites that have been visually identified to be undergoing RPS have been called jellyfish galaxies (from now on, jellyfish for short), where their stellar bodies (the jellyfish heads) remain relatively unperturbed but their gaseous disks are being stripped in the direction opposite of motion, forming the jellyfish tails \citep[e.g.,][]{Bekki2009,Ebeling2014,McPartland2016}. These jellyfish and their stripped tails are multi-wavelength objects and have been observed in the X-ray, UV, optical, and radio \citep[e.g.,][]{Gavazzi1987,Gavazzi2001,Kenney2004,Sun2006,Cortese2006,Smith2010,Jachym2017,Poggianti2019b,Ignesti2022}. However, many of these studies have focused on single or a few objects. Observers have recently pushed for systematic surveys of jellyfish galaxies, where the largest uniform samples come from the GAs Stripping Phenomena in galaxies with MUSE \citep[GASP;][54 galaxies]{Poggianti2017,Gullieuszik2020}, the OSIRIS Mapping of Emission-line Galaxies \citep[OMEGA;][70]{Chies-Santos2015,Roman-Oliveira2019}, and the LOw-Frequency ARray \citep[LOFAR;][95 in clusters and 60 in groups for 155 jellyfish in total]{Shimwell2017,Roberts2021,Roberts2021b}. The largest statistical studies of jellyfish galaxies come from \citet{Smith2022}, who use 106 jellyfish with radio continuum emission from the LoTSS survey, and from \citet{Peluso2022}, who use 131 jellyfish with information on the central ionizing mechanism.

Despite these recent efforts, unanswered questions still remain, such as: when with respect to infall and where with respect to the host does RPS begin; for how long does RPS act; did the quenched, low gas-fraction galaxies we see today go through a jellyfish phase; what determines how long RPS will take to totally remove a jellyfish's gas; how does the RPS of jellyfish galaxies compare to other satellites; where is the stripped gas being deposited, and more generally, how much cold gas do satellites bring into their hosts' halos? 

The answers to these questions can provide both insights into environmental quenching of satellites as well as important implications for the evolution of massive hosts and their surrounding halo gas in the context of the cosmic baryon cycle. While we have reached a general consensus that RPS is necessary to remove satellite cold gas and reproduce the aforementioned environmental trends, the timescales and locations of RPS and the associated satellite quenching remain highly debated. Thus, we turn to numerical simulations with temporal evolution to investigate the satellite-host interaction. Idealized simulations have been able to reproduce jellyfish by imposing an external wind, mimicking the RPS felt during infall through the CGM \citep[e.g.,][]{Tonnesen2009,Lee2020,Choi2022}. With the perspective of satellite quenching, zoom-in and full cosmological hydrodynamical galaxy simulations have studied more or less explicitly the RPS of satellites, finding a wide range of quenching timescales that broadly agree with observational inference \citep[e.g.,][]{Bahe2015,Jung2018,Wright2019,Yun2019,Oman2021,Rodriguez2022,Samuel2022b,Pallero2022,Wright2022}. However, quantitative and statistically-robust simulation predictions as to the timings and modalities of RPS are still missing. And so, to understand satellite quenching, we must first quantify the effects of perhaps its most relevant process: RPS.

In this work, we use the high-resolution, $\sim 50$~Mpc magneto-hydrodynamical simulation TNG50 from the IllustrisTNG project (TNG thereafter) to study the satellite-host interaction in a realistic, cosmological context. In particular, we aim at quantifying when, where, and for how long the RPS of cold gas occurs. We focus on cold gas as this is the source of star formation in galaxies and because its existence within the otherwise hot CGM of massive halos is a compelling open question. Moreover, we focus on jellyfish galaxies because these are satellites that, by identification and hence by construction, are surely undergoing RPS. Among its advantages, the TNG50 simulation produces thousands of galaxies and hosts ranging over 5 orders of magnitude in mass, and it naturally includes many environmental processes such as pre-processing, tidal stripping, harassment, strangulation, starvation, and RPS. The TNG simulations do not include possibly-relevant environmental processes such as viscous momentum transfer or thermal evaporation, and there is no explicit modelling of the multiphase ISM \citep[][see \citealp{Zinger2018} and \citealp{Kukstas2022} for discussions]{Cowie1977,Nulsen1982}. However, the TNG model has been shown to return satellite populations whose quenched fractions and gas content are broadly consistent with observations \citep[e.g.][]{Stevens2019,Donnari2021,Stevens2021}.

In a companion paper, \citet{Zinger2023} visually inspect TNG satellites to identify jellyfish galaxies using the citizen science Cosmological Jellyfish project hosted on Zooniverse, yielding an unprecedented number of more than 500 unique, first-infalling jellyfish galaxies in the TNG50 volume alone. In another companion paper, \citet{Goeller2023} study the star-formation activity of these jellyfish both temporally and across populations. In this paper, we employ the Monte Carlo Lagrangian tracer particles to follow the flows of gas in and out of satellite galaxies, quantifying the cold gas sources and sinks across cosmic time from when the galaxies were centrals, through their jellyfish phases, and in some cases until they have been completely stripped of all gas, existing as quenching, gas-poor satellites at $z=0$.

We begin by introducing the methods (\S~\ref{sec:meth}), namely by summarizing the TNG50 simulation (\S~\ref{sec:meth_tng}), the Cosmological Jellyfish project (\S~\ref{sec:meth_zoon}), the tracking of galaxies across cosmic time (\S~\ref{sec:meth_trees}), how we employ the tracer particles (\S~\ref{sec:meth_measurements_tracers}), and how we identify the onset and end of RPS (\S~\ref{sec:meth_measurements_timings}). We then present our main results (\S~\ref{sec:results}). We start by comparing the jellyfish galaxy population with that of the inspected and general $z=0$ satellite populations (\S~\ref{sec:results_branches}), and then comment on the origin of the jellyfish gaseous tails (\S~\ref{sec:results_jelly_tails}). After quantifying the strength of RPS post infall (\S~\ref{sec:results_RPS}) and determining a subsample of jellyfish that are devoid of cold gas at $z=0$ (\S~\ref{sec:results_end}), we answer when, where, and for how long RPS occurs (\S~\ref{sec:results_long},~\ref{sec:results_where}). We then discuss how we can generalize our jellyfish results with all $z=0$ satellites (\S~\ref{sec:disc_compare_all_satellites}), connect the the cold gas loss via RPS with satellite quenching times (\S~\ref{sec:disc_quench}, and illustrate how much and where cold gas is deposited via RPS into halos (\S~\ref{sec:disc_CGM}). We end by summarizing the main results and restating the conclusions (\S~\ref{sec:sum}).

Unless otherwise noted, all analysis including the TNG simulations adopt a $\Lambda$CDM cosmology consistent with the \citet{Planck2016} results: $\Omega_{\Lambda,0} = 0.6911, \Omega_{m,0} = \Omega_{\rm bar,0} + \Omega_{\rm dm,0} = 0.3089, \Omega_{\rm bar,0} = 0.0486, \sigma_8 = 0.8159, n_s = 0.9667, {\rm and}\ h = H_{\rm 0} / (100\, {\rm km\, s^{-1}\, Mpc^{-1}}) = 0.6774$, where $H_0$ is the Hubble parameter, and the subscript ``0" denotes that the quantity is measured today.

\section{Methods and TNG50 jellyfish galaxies} \label{sec:meth}

\subsection{The TNG50 simulation} 
\label{sec:meth_tng}

The IllustrisTNG project\footnote{\url{https://www.tng-project.org/}} \citep{Pillepich2018,Nelson2018,Naiman2018,Marinacci2018,Springel2018} consists of a series of cosmological volume $\Lambda$CDM simulations, including gravity + magneto-hydrodynamics (MHD) and a galaxy formation model \citep[see method papers for details:][]{Weinberger2017,Pillepich2018b}. Here we briefly summarize the TNG simulations. 

The TNG production simulations come in three volumes of side lengths $\sim 50, 100, {\rm and}\ 300$~comoving~Mpc, hereafter referred to as TNG50, TNG100, and TNG300 respectively. The TNG galaxy formation model was designed at the resolution of TNG100, which includes $2\times1820^3$ resolution elements with baryon mass resolution of $m_{\rm bar} = 1.4\times 10^6\, \msun$. The large volume TNG300 has $2\times 2500^3$ resolution elements with mass resolution $m_{\rm bar} = 1.1\times10^7\, \msun$. The high resolution TNG50 simulation has $2\times 2160^3$ resolution elements with mass resolution $m_{\rm bar} = 8.5\times10^4\, \msun$ \citep{Nelson2019b,Pillepich2019}. The minimum gas resolution in TNG50 at $z=0$, i.e. the smallest non-vanishing gas mass in any given galaxy, is $\approx 4\times10^{4}\, \msun$. These three simulations are publicly available in their entirety \citep{Nelson2019}. In this paper, we work exclusively with the highest-resolution run TNG50.

The TNG simulations evolve gas, cold dark matter, stars, and super massive black holes (SMBHs) within an expanding universe, based on a self-gravity + MHD framework \citep{Pakmor2011,Pakmor2013} using the \textsc{Arepo} code \citep{Springel2010}. The fluid dynamics employ a Voronoi tessellation to spatially discretize the gas. The TNG gas has a temperature floor at $10^{4}$~K, and the relationship between temperature and density for star-forming gas is determined via an effective equation of state from \citet{Springel2003}. For this analysis, we manually set the temperature of star-forming gas to $10^{3}$~K. The TNG galaxy evolution models includes the following processes: gas heating and cooling; star formation; stellar population evolution + chemical enrichment from AGB stars and type Ia + II supernovae; supernova driven outflows and winds \citep{Pillepich2018b}; formation, merging, and growth of SMBHs; and two main SMBH hole feedback modes: a thermal `quasar' mode, and a kinetic `wind' mode \citep{Weinberger2017}. The TNG simulations have reproduced many observational relations and properties across orders of magnitude in mass and spatial scales.

The group and galaxy catalogs consist of the dark matter halos and the dark matter plus baryonic galaxies. The dark matter halos are defined using the Friends-of-Friends (FoF) algorithm with a linking length $b=0.2$, run only using the dark matter particles \citep{Davis1985}. Then the baryonic components are connected to the same halos as their closest dark matter particle. Throughout this paper, we use ``FoF", ``group", ``FoF group", ``halo" synonymously. The galaxies are identified using the \subfind algorithm, which connect together all gravitationally bound particles \citep{Springel2001,Dolag2009}. We use the terms ``subhalo" and ``galaxy" synonymously even though, in general, \subfind objects may contain no stars and/or gas whatsoever. Typically albeit not always, the most massive subhalo within a halo is the ``main" or ``primary subhalo", also called the ``central galaxy"; all other subhalos within a halo are ``satellites". In all cases, we only consider subhalos of a cosmological origin as defined by the SubhaloFlag in \citet{Nelson2019}. 

\subsection{The Cosmological Jellyfish project on Zooniverse} \label{sec:meth_zoon}

In this paper, we study jellyfish galaxies from the TNG50 simulation and identify them based on the classification of the Zooniverse Cosmological Jellyfish project\footnote{\url{https://www.zooniverse.org/projects/apillepich/cosmological-jellyfish}}. The Zooniverse Cosmological Jellyfish project presented images of TNG50 satellite galaxies -- in addition to TNG100 galaxies, not studied here -- on the Zooniverse platform for classification by citizen scientists. Here several thousand volunteers underwent a training session and classified whether the given galaxy resembles a jellyfish or not \citep{Zinger2023}. 

Following the pilot project that visually classified a subset of TNG100 satellites \citep{Yun2019}, the term ``jellyfish galaxy" was associated with a satellite with a visually identifiable signature of RPS in the form of asymmetric gas distributions in one direction. The visual inspection is based on images of gas column density -- i.e. all gas irrespective of phase, temperature, etc. -- with stellar mass contours, projected in random orientations in a field of view of 40 times the 3D stellar half mass radius $R_{\rm half,\star}$. Each image was classified by at least 20 inspectors (trained volunteers) whose proficiency was measured when tallying the votes. A galaxy image received a score between 0 and 1 based on these votes, whereby we employ a threshold of 0.8 and above to identify jellyfish galaxies, as recommended by \citet{Zinger2023}.

Galaxies meeting the following criteria had their images posted for inspection for the Zooniverse project:
\begin{itemize}
    \item non central, i.e. satellite;
    \item of cosmological origin, as defined by the SubhaloFlag in \citet{Nelson2019};
    \item $\mstarsat \equiv M_{\star}^{\rm sat}(<2\times R_{\rm half,\star}) > 10^{8.3}$;
    \item $f_{\rm gas} \equiv M_{\rm gas}^{\rm sat} / \mstarsat > 0.01$, where $M_{\rm gas}^{\rm sat}$ is the satellite's total (i.e. gravitationally-bound) gas mass.
\end{itemize}

All galaxies satisfying the above criteria were inspected at each available snapshot since $z=0.5$ (every $\sim150$~Myr in cosmic time; snapshots 99-67), and at redshifts $0.7, 1.0, 1.5,\ \text{and}\ 2.0$ (every $\sim1$~Gyr in cosmic time; snapshots 59, 50, 40, and 33).

According to the results of the Zooniverse Cosmological Jellyfish project for TNG50, 4,144 of the total 53,610 (7.7~per~cent) galaxy images are jellyfish. See \citet{Zinger2023} for more details on the Zooniverse Cosmological Jellyfish project and related results for both TNG50 and TNG100.

\subsection{Tracking galaxies along the merger trees} \label{sec:meth_trees}

Based on the selection for the Zooniverse Cosmological Jellyfish project, frequently an individual galaxy was inspected multiple times at different points in time along its evolutionary track.

In this paper, we connect the galaxies that were inspected at multiple times using \sublinkgal \citep{Rodriguez-Gomez2015}. Briefly, \sublinkgal constructs the merger trees at the subhalo level by searching for descendant candidates with common stellar particles and star-forming gas cells. Then \sublinkgal chooses the descendant by ranking all candidates with a merit function that takes into account the binding energy of each particle/cell, and choosing the candidate with the highest score as the descendant.

In this paper, we chiefly work with and follow the unique evolutionary tracks of galaxies, branches, inspected in the Cosmological Jellyfish project. In total, there are 5,023 unique galaxy branches in TNG50 among the inspected images. The analysis of these satellite galaxy populations along their evolutionary tracks requires following the merger tree branches both of the individual galaxies and their (sometimes temporary) hosts. We give results on this in \S\ref{sec:results_branches} and more details in Appendix~\ref{app:tracking}.

\subsection{Galaxy sample selection of this analysis} \label{sec:meth_sample}

With respect to the Zooniverse Cosmological Jellyfish project, we apply additional selection criteria to be able to start from a sample of satellites defined at $z=0$ that does not include backsplash and pre-processed galaxies. Please see Appendix~\ref{app:tracking} for details regarding how we classify the galaxies as backsplash and/or pre-processed.

Of the 5,023 inspected galaxy branches in TNG50, we apply the following sample selection criteria. At each criterion, we list the number of remaining branches in the simulation, and the number excised by this criterion in parentheses\footnote{The number excised is the number from the previous criterion. For example, criterion (ii) excises $x$ branches from the $y$ branches remaining after applying criterion (i). This now leaves $y-x$ branches after applying criterion (ii).}.
\begin{enumerate}
    \item The galaxy must survive until the end of the simulation at $z=0$. That is, the main descendant branch must track the subhalo until snapshot 99: 3,018 (2,005 excised).
    \item There must be at least one snapshot since $z \leq 0.5$ when the galaxy was inspected in the Zooniverse project (and therefore meeting the criteria outlined in \S\ref{sec:meth_zoon}): 2,398 (620). 
    \item The galaxy must be a satellite galaxy at $z=0$, i.e., not a backsplash galaxy at snapshot 99: 2,062 (336). 
    \item The galaxy must not have been pre-processed by a host group other than its $z=0$ host: 1,610 (452).
    \item The galaxy must have a well defined infall time (must have been a central galaxy for at least one snapshot before becoming a satellite): 1,543 (67).
\end{enumerate}
Thus our total number of cleaned, first-infalling inspected branches in TNG50 is 1,543. Of these branches, we separate them into those that have at least one jellyfish classification since $z = 0.5$, called ``jellyfish" branches, and those without a jellyfish classification since then, called ``non-jellyfish" branches. The numbers of jellyfish and of non-jellyfish branches in TNG50 are 512 (33~per~cent) and 1,031 (67~per~cent), respectively\footnote{In TNG50, there are 8 cleaned, inspected branches that have a jellyfish classification before $z=0.5$ but not afterwards. We exclude these galaxies from the jellyfish sample.} (see \S\ref{sec:results_branches} for additional results). We note that at the time of infall all inspected branches (jellyfish and non-jellyfish) are star-forming; see \S~\ref{sec:disc_quench} for a discussion regarding the quenching times and \citet{Goeller2023} for details on the star-forming properties of these galaxies.

\subsection{On cold gas, infall time, tracer particles, and measuring ram pressure stripping} \label{sec:meth_measurements}

In this work, we study the gravitationally-bound cold gas of TNG50 satellite galaxies: by cold gas, throughout this paper, we mean gas with a temperature $T_{\rm ColdGas} \leq 10^{4.5}$~K (including star-forming gas; see \S~\ref{sec:meth_tng} for more details).

Throughout this paper, we define infall as the first time in cosmic history that a galaxy becomes a satellite member of its $z=0$ FoF host, irrespective of distance.

\subsubsection{Following the gas with tracer particles}
\label{sec:meth_measurements_tracers}
As TNG50 is based on a moving-mesh code to follow the evolution of the underlying fluid field, we must employ the Monte-Carlo-Lagrangian tracer particles to follow the history and evolution of individual gas parcels \citep{Genel2013,Nelson2013}. Briefly, {\sc arepo} treats the gas as a fluid field through a Voronoi mesh. There is no innate method to follow the flow of matter between the mesh elements and across time. Thus the tracers are introduced, acting as test particles within the fluid. TNG50 was run with one tracer per gas cell at the initial conditions. The tracers have a constant identifying number (ID) throughout the simulation, and at each snapshot each tracer has exactly one baryonic parent resolution element: a gas cell, a stellar or wind particle, or a SMBH. This means that any given tracer represents $8.5\times10^{4}\, \msun$ of baryonic mass with the properties of its parent. For example, if a single tracer has a gas parent at one time and a star parent at the next time, then the tracer represents $8.5\times10^{4}\, \msun$ of gas mass being converted into stars. In this way, one can track the flow of matter by following a given tracer and its parent's properties across cosmic time. In TNG50 the parents of the tracers are output at each snapshot, describing the exchange of parcels of baryonic material across resolution elements at time intervals of $\sim 150$~Myr. As the tracer particles are Monte Carlo in nature, we make only statistical statements about the behavior of thousands to millions of tracers.

In practice, at each snapshot and for each galaxy of interest, we find all tracers whose parents are bound, cold gas cells. While not every gas cell necessarily has an associated tracer and some gas cells may have multiple child tracers, the total tracer cold gas mass (total number of tracers times $m_{\rm bar} = 8.5\times10^{4}\, \msun$) agrees with the total amount of cold gas mass measured by gas cells (see \S~\ref{sec:results_RPS} for more details and an example). Then we follow the tracers and their parents across snapshots in order to measure the cold gas mass that is stripped or launched in an outflow, becomes hot, participates in star formation, transforms into a wind particle, and gets accreted into a SMBH.

We proceed as follows, on a galaxy by galaxy basis along its main descendant branch (MDB). Starting from the first snapshot that the galaxy is identified in the merger trees, we find all tracers whose parents are bound, cold gas cells of this galaxy. Then at the next snapshot for the galaxy along its MDB, we find which tracers belong to one of the  following mutually exclusive and completely exhaustive groups:
\begin{enumerate}
\item are recorded in both snapshots: bound, cold gas that remains bound, cold;
\item are recorded in the current snapshot but not in the previous one: currently bound, cold gas that previously was either not bound or not cold;
\item are recorded in the previous snapshot but not the current one: previously bound, cold gas that no longer is;
\end{enumerate}
Potential physical origins of tracers in group (ii) include inflows, cooling, stellar mass return, or wind re-coupling. The group (iii) tracers could either a) go from cold gas cells into one of the following: star particles (star formation denoted SF, or SFR for star formation rate); SMBH sink particles (i.e. SMBH accretion); bound, warm/hot gas cells (heating); or b) be no longer bound gas cells (stripping + outflows). We denote the latter ``RPS+outflows'' and will be focusing on this quantity throughout the paper. We include tracers whose parents become unbound and hot in the same time step in this category. We note that tidal stripping may be included in RPS+outflows, although visual inspection shows that ram-pressure stripping is the dominant mechanism of jellyfish galaxies,  and a majority of galaxies do not reach host-centric distances $\lesssim 0.2\rvir$. Moreover, the Zooniverse inspectors were specifically asked {\it not} to classify an image as a jellyfish if there was a close companion or gaseous tails were visible on both sides of the galaxy \citep{Zinger2023}.

\subsubsection{Identifying the onset and end of ram pressure stripping} \label{sec:meth_measurements_timings}

Throughout our analysis, prior to infall (host FoF membership; see above), we assume that the RPS + outflows category is dominated by outflows, namely outflows driven from stellar- and/or SMBH-feedback. As we further justify in \S~\ref{sec:results_RPS}, for most jellyfish the amount of outflows before infall is approximately constant. Immediately after infall, there is commonly an increase in the RPS + Outflows category, indicating that another physical process has become present, namely RPS. Moreover between infall and pericenter, many satellites experience bursts of star formation and/or AGN accretion, which has also been seen in observations, reproduced by simulations, and thought to be caused by ram pressure compressing the ISM gas \citep{Gavazzi1987,Bahe2015,Mistani2016,Zoldan2017,Vulcani2018,Roberts2020,Grishin2021,Peluso2022,Garling2022,Goeller2023}. These bursts of star formation and/or AGN accretion would in turn induce turbulence in the ISM and drive outflows, which then facilitate ram-pressure stripping \citep[e.g.,][]{Bahe2015}. Attempting to distinguish the relative contributions from outflows and ram pressure becomes a chicken-and-egg problem. Thus, we consider the time of infall to be the onset of RPS, and after infall relabel the quantity ``RPS+Outflows" as ``RPS". We note that we have estimated the onset of RPS using two alternative methods, and find that for most jellyfish the difference between the various methods is $\lesssim 450$~Myr ($\lesssim 3$~snapshots): see Appendix~\ref{app:tau} and Fig.~\ref{fig:tau0_contours} for more details.

The end of ram-pressure stripping is either when the galaxy's cold gas mass falls below our resolution limit (namely below  $\approx4\times10^4\,\msun$ i.e. $f_{\rm gas} \lesssim 5\times10^{-4}$ for a galaxy at our minimum stellar mass of $\mstarsat = 10^{8.3}\, \msun$), or the end of simulation at $z=0$. In our sample, 259/512 ($\approx 50$~per~cent) galaxies lose all their cold gas at or before $z=0$.

We denote the onset of RPS as the infall time $\tau_0$ and its end as $\tau_{100}$ (when 100~per~cent of the RPS has occurred), so that the difference between these two times returns in principle the maximum time span over which RPS has acted on any given galaxy: 
\begin{equation} \label{eqn:RPS}
\tau_{\rm RPS} = \tau_{100} - \tau_{0},
\end{equation}
where $\tau_{100}, \tau_{0}$ are the ages of the universe at the given points. This RPS time span is the longest duration over which RPS has acted  for the galaxies that have lost their cold gas prior to $z=0$. On the other hand, for those satellites that still have some gas today, the above-defined timescale of RPS is likely a lower estimate, while we speculate that these galaxies would continue being stripped in the future. See \S\ref{sec:results_end} for differences between these two ending states.

Throughout the paper, we will compare the times of RPS with estimates of the quenching time, i.e. of the most recent and last time that a galaxy has fallen 1~dex below the star-forming main sequence (SFMS) for its mass and redshift, as per definitions of \citet{Pillepich2019} and catalogs from \citet{Joshi2021,Donnari2021}.

\section{Results} \label{sec:results}

\subsection{TNG jellyfish galaxies across their unique branches} \label{sec:results_branches}

\begin{figure*}
    \includegraphics[]{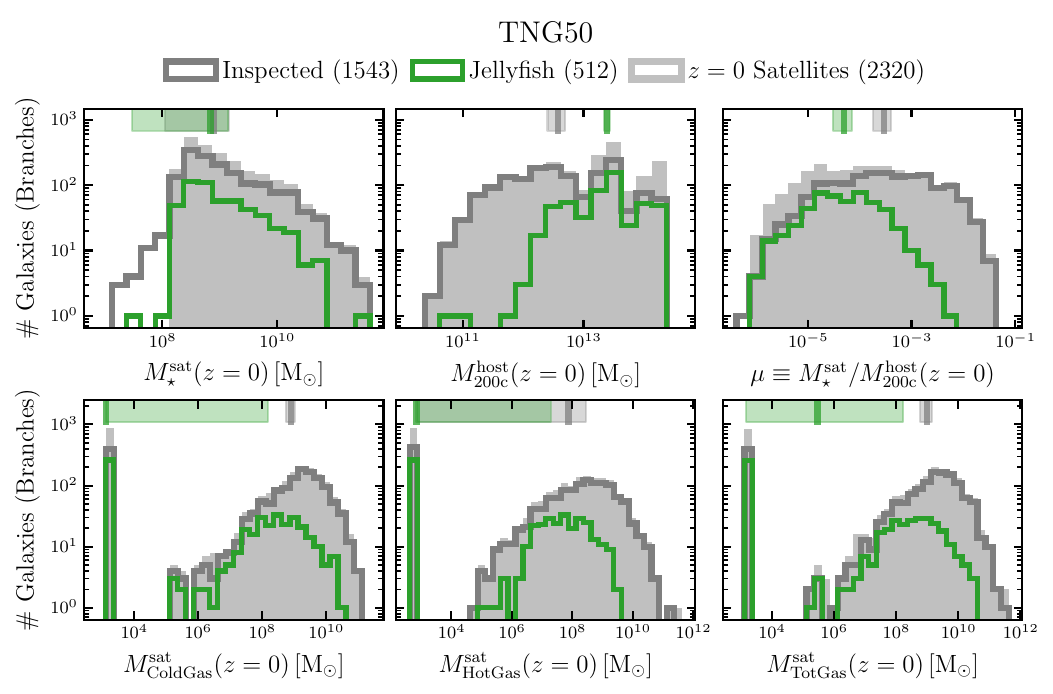}
    \caption{
    {\bf Selection of TNG50 galaxies studied in this work and the abundance of jellyfish, along their unique branches.}
    The Inspected sample (dark gray) includes a subset of all satellite branches from TNG50, selected for the identification of galaxies with clear signatures of RPS: this chiefly excludes satellites with $\mstarsat < 10^{8.3}\, \msun$ and less than 1~per~cent of gas mass fraction at the time of inspection, as well as pre-processed and backsplash galaxies. The Jellyfish sample (green) also requires at least one jellyfish-classified snapshot at $z \leq 0.5$. See \S\ref{sec:meth_zoon} and \ref{sec:meth_sample} for more details. The medians and $1\sigma$ errors of the Inspected and Jellyfish galaxy distributions are marked by the hash marks and shaded regions on the top $x$-axis. For comparison, we show all TNG50 $z=0$ satellites with $\mstarsat > 10^{8.3}\, \msun$ (light gray). For the gas properties in the bottom row, the galaxies with gas masses below our resolution limit are placed manually at $\sim 10^3\, \msun$. Cold gas has temperatures $\leq 10^{4.5}$~K; hot gas has temperatures $ > 10^{4.5}$~K.
   }
    \label{fig:TNG50_selection_hists}
\end{figure*}

According to the Cosmological Jellyfish project on Zooniverse, 4,144 of the 53,610 images from TNG50 are jellyfish galaxies \citep[7.7~per~cent;][]{Zinger2023}. Using the merger trees to identify when the same galaxies were imaged at multiple points in cosmic time in TNG50 and applying our selection criteria (\S\ref{sec:meth_sample}), we now focus on our sample of 512 first-infalling unique jellyfish galaxies, among 1,543 unique, inspected branches (33~per~cent). 

Fig.~\ref{fig:TNG50_selection_hists} shows our selection of Jellyfish (green histograms) and Inspected (dark gray histograms) satellites at $z=0$. We now quote numbers in terms of unique branches such that Fig.~\ref{fig:TNG50_selection_hists} is the branch- or merger tree-based counterpart of similar histograms in \citet[][see their fig.~2]{Zinger2023}. In each of the panels, we include the medians and $1\sigma$ errors (hashes and shaded regions on the top $x$-axis) for the Inspected and Jellyfish samples. We note that for each of the distributions (except $\mstarsat(z=0)$, see text below), the 2-sample Kolmogorov-Smirnov (KS) and Anderson-Darling (AD) tests suggest at $\geq95$~per~cent confidence that the Inspected and Jellyfish samples were not drawn from the same parent distribution, i.e. that the two samples are significantly different. We include for comparison the general population of $z=0$ satellites with $\mstarsat(z=0) \geq 10^{8.3}\, \msun$, which is generally similar to the Zooniverse inspected sample, except that the general $z=0$ satellite population includes pre-processed satellites. See \S~\ref{sec:disc_compare_all_satellites} for a more detailed discussion on how representative the jellyfish sample is compared to all $z=0$ satellites above stellar mass.

Firstly, Fig.~\ref{fig:TNG50_selection_hists} shows, thanks to TNG50, that we can study satellite galaxies, and hence jellyfish and RPS, in a rather extended range of stellar masses and host masses. Namely we study satellites with stellar masses $\sim10^{8-12}\, \msun$ orbiting in hosts with total masses $\sim10^{10.5-14.3}\, \msun$ at $z=0$. However we cannot make statements about satellites in the most massive clusters $\mvir \sim 10^{15}\, \msun$.

\begin{figure*}
    \includegraphics[width=0.96\textwidth]{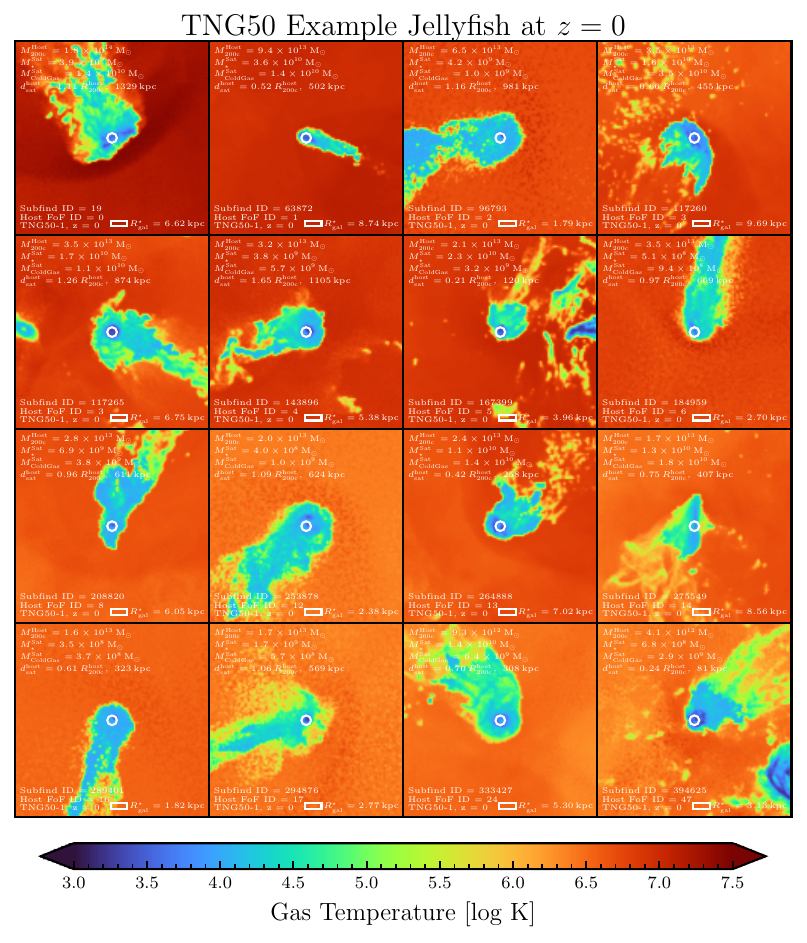}
    \caption{
    {\bf The coldness of the ram-pressure stripped gas in TNG50 jellyfish galaxies.}
    We show gas temperature maps of 16 TNG50 jellyfish galaxies, randomly chosen at $z=0$. Each image is $(40\times R_{\rm half,\star})^3$ in size and depth, with $100\times 100$ pixels ($\sim$~kpc sized pixels) in the same orientation as the jellyfish were posted to Zooniverse (i.e., random and along the $z$-axis). Here, we measure the mass-weighted-average temperature map of all (FoF i.e. ambient) gas within the cube, and overplot the jellyfish (i.e. gravitationally-bound) gas. The white circle shows the galaxy stellar radius ($\rgal = 2\times R_{\rm half,\star}$), and information about the jellyfish galaxy and its host are in the top- and bottom- left corners. Star-forming gas is placed at the nominal temperature of $10^{3}~$K, so all dark blue locations represent active star-forming regions. The gas in the jellyfish tails is typically and on average cold-cool $\sim10^{4-5}$~K.
    }
\label{fig:TNG50_temperature_poster}
\end{figure*}

\begin{figure*}
    \includegraphics[width=0.96\textwidth]{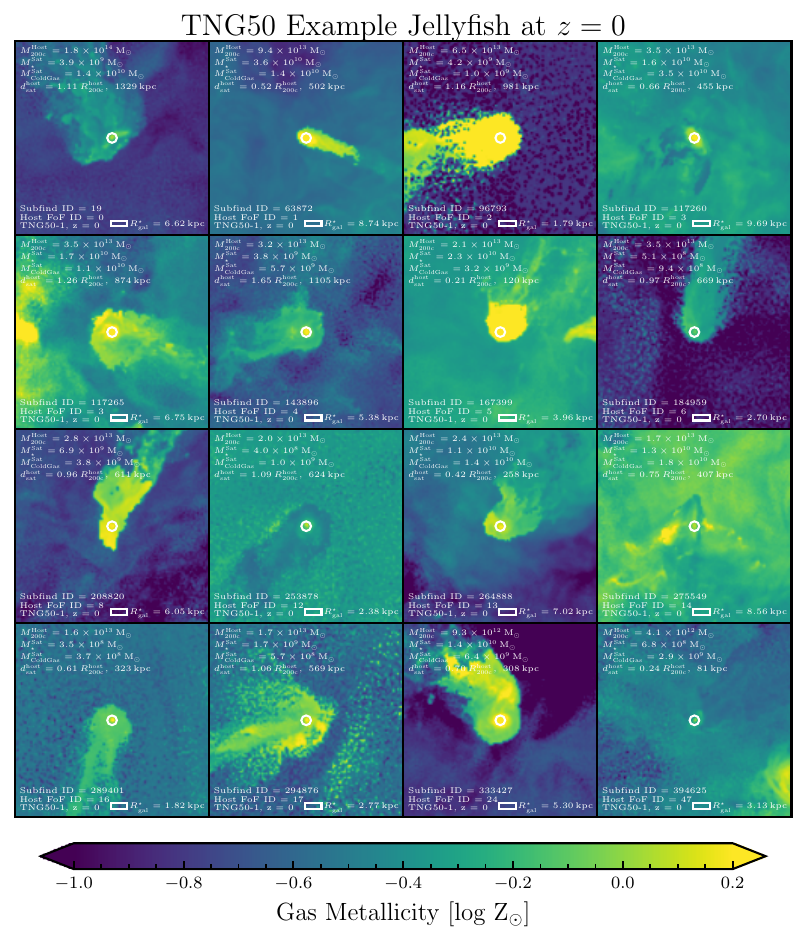}
    \caption{
    {\bf The metallicity of the ram-pressure stripped gas in TNG50 jellyfish galaxies.} Similar to Fig.~\ref{fig:TNG50_temperature_poster} but here showing the mass-weighted gas metallicity rather than the temperature. The tails of jellyfish are as enriched as the main body of the satellites they stem from, but depending on the host, the tails may or may not be more enriched than the ambient gas.}
\label{fig:TNG50_metallicity_poster}
\end{figure*}

As shown in the top left panel, jellyfish galaxies (green) tend towards lower stellar masses $\mstarsat(z=0)$ compared to the inspected galaxies (dark gray), and especially to the non-jellyfish galaxies that have been inspected (not shown, but would be dark gray minus green). Since the stellar body is the primary foil to RPS, providing the gravitational binding energy for the gas to remain in the galaxy, galaxies with a weaker restoring force are naturally more susceptible to RPS, in line with other studies of TNG jellyfish \citep{Yun2019,Zinger2023}. Because we only inspect galaxies with $\mstarsat > 10^{8.3}\, \msun$ at the time of inspection, we see a decrease in the number of galaxies at lower masses. This inspection criterion is only at the snapshot of inspection, so galaxies that later lose stellar mass due to either tidal stripping or stellar mass return may have stellar masses below this lower limit. The fact that only 2/512 (0.39~per~cent) of jellyfish branches compared to 26/1,031 (2.5~per~cent) of non-jellyfish branches have stellar masses below the inspection criterion suggests that we are able to separate galaxies undergoing tidal vs ram-pressure stripping. At the high-mass end, $\mstarsat \gtrsim 10^{10.5}\, \msun$ there are only a few jellyfish galaxies. We speculate that this is a combination of two effects: more massive satellites in hosts of this mass range better retain their cold gas against stripping; at these stellar masses, the TNG kinetic mode of SMBHs expels much of the galaxy's gas \citep[e.g.][]{Terrazas2020,Zinger2020}, often at infall and before the peak effectiveness of RPS. While the AD test suggests confidence at the $\approx95$~per~cent level that the two distributions are distinct, the KS test suggests only $\approx$85~per~cent confidence, and the medians of the two distributions are not significantly different.

In the top middle panel, we see that jellyfish typically live in more massive hosts, and almost all inspected galaxies in massive hosts $\mvirhost \gtrsim 10^{13}\, \msun$ have been classified at some point since $z=0.5$ as jellyfish. The number of satellite galaxies increases with the host halo mass due to hierarchical structure formation. With increasing host mass the gravitational potential well deepens, which in turn leads to both better retention of stellar- and SMBH-driven outflows from the central and more cosmological gas accretion from the large scale structure. These effects generally lead to a denser CGM or ICM. Moreover, deeper potential wells increase the infall velocities of satellite galaxies, sometimes even to supersonic speeds \citep{Yun2019}. The denser ambient medium and the increased relative velocity both increase the strength of ram-pressure stripping \citep[e.g.][]{Yun2019}. However in the past five billion years, MW-mass halos $\mvirhost \sim 10^{12}\, \msun$ have also hosted a number of jellyfish galaxies. 

In the top right panel Fig.~\ref{fig:TNG50_selection_hists}, by combining the effects of satellite stellar mass and host mass, we see that jellyfish galaxies typically have small mass ratios $\mu \equiv \mstarsat / \mvirhost$, and nearly every inspected galaxy with a mass ratio $\mu \lesssim 10^{-4}$ is a jellyfish. 

The satellite stellar mass distribution of the inspected galaxies (dark gray) is slightly below but quite similar to that of the $z=0$ satellites (light gray) for stellar masses $\mstarsat \sim 10^{8.3-10.5}\, \msun$ (top left), and the distributions are nearly identical for masses $\mstarsat \sim 10^{10.5-11.8}\, \msun$. Compared to the $z=0$ satellites, the inspected galaxies have an under-population of high mass hosts $\mvirhost\sim10^{13.5-14.3}\, \msun$ (top middle) and low mass ratios $\mu \lesssim 10^{-4}$ (top right). We speculate that many of these $z=0$ satellites are pre-processed and therefore have been excluded from this analysis, but they may also have had too low of gas masses and their fractions to be inspected (bottom panels).

In the bottom panels of Fig.~\ref{fig:TNG50_selection_hists}, we see that jellyfish galaxies typically exhibit, at $z=0$ lower amounts of gravitationally-bound cold gas $\mcgassat$ with temperatures $T_{\rm ColdGas} \leq 10^{4.5}$~K (or star-forming; bottom left), hot gas $M_{\rm HotGas}^{\rm sat}$ with temperatures $T_{\rm HotGas} > 10^{4.5}$~K (bottom middle) and total gas $M_{\rm TotGas}^{\rm sat}$ (bottom right) compared to the inspected branches. A larger fraction of jellyfish ($\approx50$~per~cent) compared to non-jellyfish ($\approx12$~per~cent) have gas masses below our resolution limit, plotted here at $M_{\rm Gas}^{\rm sat}\sim 10^{3}\, \msun$. We have explicitly checked that the non-jellyfish inspected galaxies with large $z=0$ gas reservoirs are typically late-infallers and have higher mass ratios, causing weaker ram-pressure stripping. Conversely, the non-jellyfish inspected satellites without any gas at $z=0$ are typically early-infallers, namely they joined their $z=0$ hosts when galaxies were inspected only every $\sim1$~Gyr, compared to every $\sim150$~Myr after $z = 0.5$. Additionally, there are a few cases of massive galaxy mergers where the FoF-identified central galaxy switches between the two galaxies; this means that these quasi-central galaxies meet the inspection criteria but are not truly classical satellites. 

\subsection{Jellyfish tails stem from the stripped, cold ISM} \label{sec:results_jelly_tails}

\begin{figure*}
    \includegraphics[]{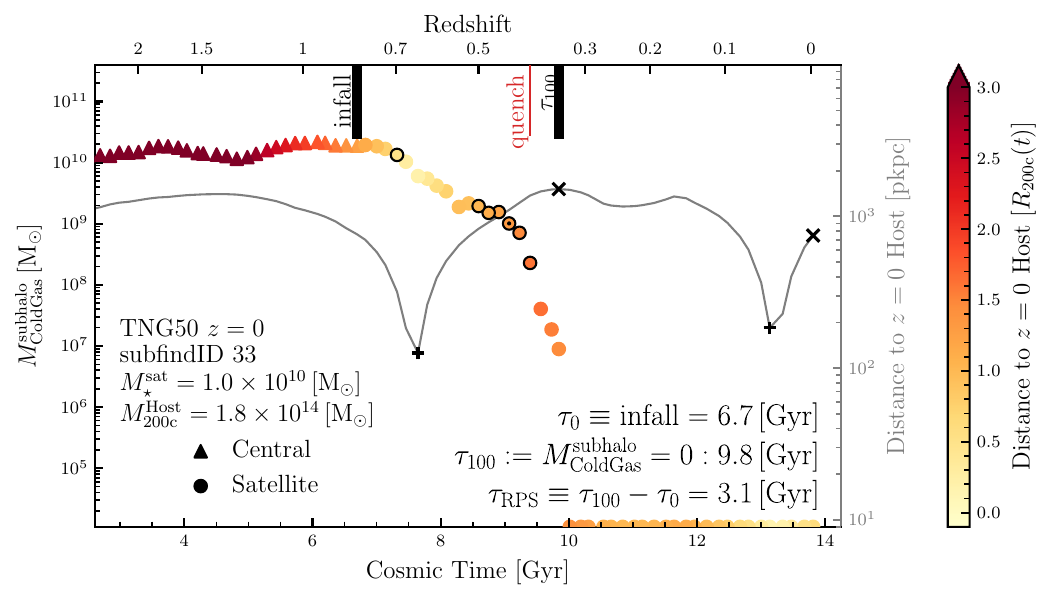}
    \caption{
    {\bf Time evolution of the total gravitationally-bound cold gas $\mcgassub$ of a TNG50 jellyfish galaxy.}
    The marker style denotes the FoF membership of the galaxy, as either a central (triangle) or satellite (circle), whereas the color indicates the distance to the $z=0$ host in normalized units of [$\rvir(t)$]. The host-centric distance in physical units (thin gray curve) uses the right y-axis, and we mark the pericentric and apocentric passages with black ``+" and ``x" symbols respectively. The snapshots when this galaxy has been visually inspected are outlined with a thick black circle and the snapshot(s) when it has been classified as a jellyfish are indicated with a central black dot. We place by hand the times after $\tau_{100}$ when the cold gas mass $\mcgassub$ is below our resolution limit ($\lesssim 4\times10^4\, \msun$) at the lower y-limit (along the bottom x-axis). The thick black ticks denote the onset of RPS as the infall time ($\tau_0$) and the end of RPS ($\tau_{100}$), in this case when $\mcgassub(t) = 0$. The red tick marks when the galaxy quenches, defined as when the galaxy falls at least one~dex below the star-forming main sequence for the last time.
    }
    \label{fig:SCGM_evolution}
\end{figure*}

In this work, we study the ram-pressure stripping of cold gas because the long-lived jellyfish tails originate mostly from the cold ISM of satellite galaxies. We provide arguments for this as follows.

Firstly in Fig.~\ref{fig:TNG50_temperature_poster} we show the gas temperature maps of 16 TNG50 jellyfish at $z=0$. Each image is $(40\times R_{\rm half,\star})^3$ in size and depth, with $100\times 100$ pixels ($\sim$~kpc sized pixels) in the same orientation as the jellyfish were posted to Zooniverse (i.e., random and along the $z$-axis). We measure the mass-weighted-average temperature map of all (FoF i.e. ambient) gas within the cube, and overplot the jellyfish (i.e. gravitationally-bound) gas. In each image, the jellyfish tails' temperature matches, or is at a similar temperature of, the ISM gas, which we roughly denote as the gas enclosed by the white circles of radius $2\times R_{\rm half,\star}$. In some cases, a bow shock is also present, which appears as a stark contrast in temperature in the opposite direction of the tails \citep[e.g., top left; see also other manifestations of bow shocks in front of TNG100 jellyfish galaxies in fig.~10 of][]{Yun2019}. 

Fig.~\ref{fig:TNG50_metallicity_poster} showcases the metallicity maps for the same 16 TNG50 jellyfish galaxies. Generally, the metallicity of the jellyfish tails is similar to that of the main body of the galaxy (the jellyfish head). Unlike the temperature, the metallicity of the background halo gas is not always so distinct from the jellyfish (e.g., bottom right), as it depends on the satellite-to-host mass ratio.

\begin{figure*}
    \includegraphics[]{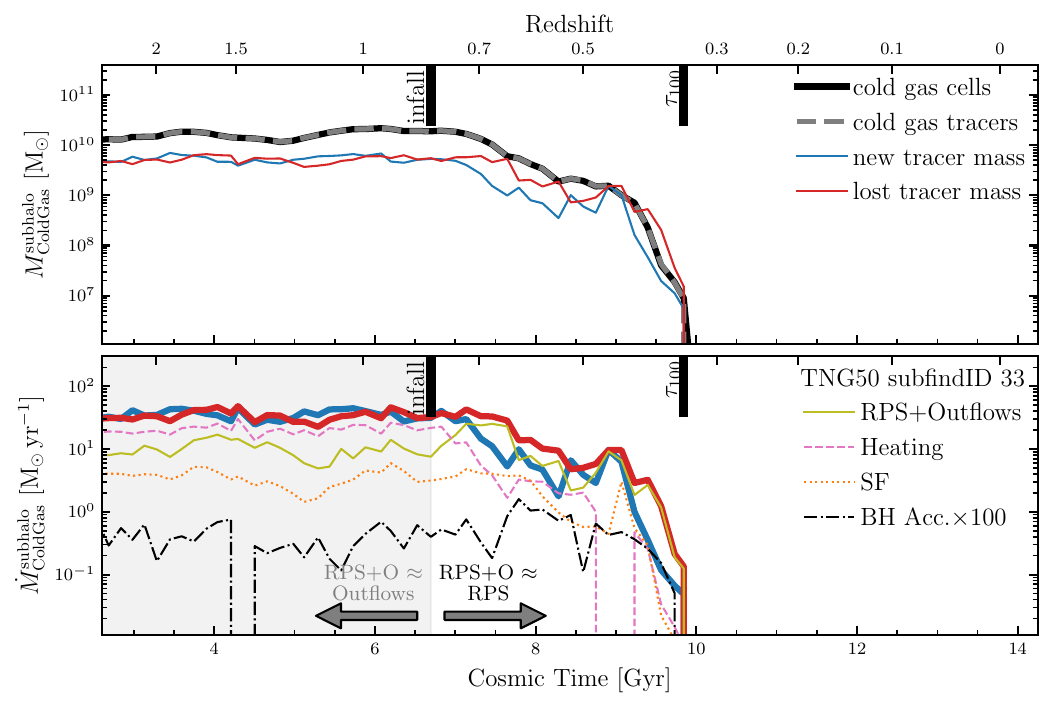}
    \caption{
    {\bf Time evolution of the tracer quantities for an example TNG50 jellyfish galaxy, the same galaxy as in Fig.~\ref{fig:SCGM_evolution}.}
    In the top panel, we see that the total cold gas mass (thick black curve) and tracer mass with cold gas parents (gray dashed curve) agree. The thin blue curve shows the the total new cold gas, the tracer mass whose parents are now cold gas cells but previously were not, such as cooling or inflowing gas. The thin red curve shows the opposite, and we further separate the various physical mechanisms of cold gas loss in the bottom panel, now normalized by the time between snapshots. The contribution to net lost tracers from star formation (orange dotted) is small at all times, while heating (pink dashed) is dominant only while the galaxy is a central. The cold gas mass lost via SMBH accretion (black dot-dashed) is shown here multiplied by 100 and negligible at all times. When the galaxy becomes a satellite at infall (black tick mark, cosmic time 6.7 Gyr, $z\sim0.8$), RPS (olive) becomes the dominant source of cold gas loss. Throughout this paper and following the inspection of the evolutionary tracks of all selected galaxies, we assume that before infall, RPS + Outflows is dominated by outflows, and that after infall RPS is dominant.
    }
    \label{fig:tracer_evolution}
\end{figure*}

These images exemplify that, at the time when a RPS tail is identifiable in gas column density, the physical properties of the gas in the tails are similar to those in the ISM in the main body of the satellite galaxy undergoing RPS. The tail gas is cold and is typically as metal enriched as the jellyfish head.

Furthermore, we have checked that, at the time of infall, $\approx75~(60)$~per~cent of the gravitationally-bound gas mass is cold for jellyfish galaxies with stellar masses at infall of $\mstarsat(\tau_0) = 10^{8-9}~(10^{9-10})~\msun$.

As a note, this ISM-origin of the RPS'ed gas does not preclude the jellyfish tails to reveal themselves across a wide range of wavelengths (see \S~{\ref{sec:intro} for references). Namely, although the bulk of the tail gas is cold or cool according to TNG50, it can also manifest itself in e.g. soft X-ray \cite[see fig.~12 from][for a mock 100~ks exposure from the Line Emission Mapper for an example TNG100 jellyfish galaxy in the soft X-ray continuum and at the OVII f line]{Kraft2022}.

\subsection{The majority of the cold gas loss after infall is due to ram-pressure stripping} \label{sec:results_RPS}

According to TNG50, RPS is the dominant source of cold gas loss after infall for jellyfish galaxies. This is somewhat to be expected, given the jellyfish nature of the selected galaxies under inspection. However, we have demonstrated this for all 512 TNG50 jellyfish galaxies, using the tracer particle analysis described in Section~\ref{sec:meth_measurements_tracers}. We showcase this result with one example galaxy below.

Fig.~\ref{fig:SCGM_evolution} shows the time evolution of the gravitationally-bound cold gas mass $\mcgassub$ for one example TNG50 jellyfish galaxy. Prior to infall (at cosmic times $\lesssim 6.5$~Gyr) and at large distances ($\gtrsim 2\rvirhost \sim 10^3$~kpc), the cold gas associated to the galaxy is approximately constant. After infall, $\mcgassub$ decreases significantly through the first pericentric passage until the satellite has effectively no cold gas remaining, which we denote as $\tau_{100}$ (see \S\ref{sec:meth_measurements_timings} for more details). The galaxy quenches its star formation for the last time shortly before $\tau_{100}$, at $\approx3$~Gyr after infall. 

Fig.~\ref{fig:tracer_evolution} graphs the evolution of cold gas mass and the associated tracers for the same galaxy as in Fig.~\ref{fig:SCGM_evolution}. In the top panel, $\mcgassub$ measured using the gas cell data (thick black curve) is identical to that in Fig.~\ref{fig:SCGM_evolution}; moreover, $\mcgassub$ measured using the tracers (dashed gray curve; the number of tracers with cold gas parents times the baryonic mass resolution) closely matches the cold gas mass measured using \subfind at all times. This affirms that the tracers robustly measure the cold gas mass (see \S~\ref{sec:meth_measurements} and Appendix~\ref{app:tracking} for more details). 

In the top panel of Fig.~\ref{fig:tracer_evolution}, while the galaxy is a central before infall at cosmic times $\lesssim 6.7$~Gyr, the net new (thin blue curve) and lost (thin red curve) cold gas tracers roughly balance each other, leading to the approximately constant total $\mcgassub$. This likely reflects a quasi-equilibrium galactic fountain scenario, where inflows and outflows approximately cancel out to yield a constant $\mcgassat$, at least for the depicted galaxy. At infall, there is an immediate drop in new cold gas -- the cold gas that the galaxy acquired via cold gas inflows or gas cooling -- which qualitatively agrees with the results from the EAGLE simulation \citep{Wright2019,Wright2022}. The lost cold gas mass remains approximately constant for $\sim1$~Gyr after infall before eventually declining. After infall, the lost cold gas is always similar to or higher than the new cold gas, leading to the net decline in cold gas mass until $\mcgassub < 4\times10^{4}\, \msun$. However it is interesting that the new cold gas remains nonzero for Gyrs after infall, including during the pericenter passage.

In the bottom panel of Fig.~\ref{fig:tracer_evolution}, we show again the net new (thick blue curve) and lost (thick red curve) tracers of cold gas, now shown as cold gas mass rates normalized by the time between snapshots. Further, we split the lost tracers into the various sinks of RPS+outflows (solid olive), gas heating (dashed pink), star-formation (SF; dotted orange) and SMBH accretion multiplied by 100 (BH acc.; black dot-dashed). See \S~\ref{sec:meth_measurements_tracers} for additional technical inputs.

Before infall, gas heating is the dominant mechanism of cold gas loss, followed by RPS+outflows and SF. During this time, the shapes of the RPS+outflows, SF and SMBH accretion$\times$100 are quite similar, suggesting that SF and/or SMBH accretion are the primary drivers of outflows for this galaxy. For the first $\sim$Gyr after infall, the SF remains roughly constant while the RPS+outflows increases, confirming the onset of RPS. Moreover there is a simultaneous net gas loss, translating into an increase in the ``efficiency" of RPS+outflows and SF, where efficiency here denotes RPS+outflows or SF normalized by $\mcgassat$. During this period, the cold gas lost via heating also decreases significantly. SMBH accretion is the least dominant cold gas sink at all times, at least for this galaxy. This galaxy has in fact experienced little to no kinetic AGN feedback, though in general 45 of the 512 ($\approx 9$~per~cent) of jellyfish galaxies have $\mstarsat(z=0) > 10^{10}\, \msun$ and have experienced kinetic AGN feedback. Through pericenter until the jellyfish has a gas mass below our resolution limit, RPS+outflows remains the dominant physical mechanism of cold gas removal.

As discussed and anticipated in \S\ref{sec:meth_measurements_tracers}, outflows and RPS are closely intertwined, for example as outflowing gas is less gravitationlly bound and therefore more susceptible to RPS. We hence avoid distinguishing between cold gas that is lost (and becomes unbound) because of RPS or because of a combination of RPS and high velocities, and we conclude that that RPS (+outflows) is the dominant source of cold gas loss after infall for jellyfish galaxies in TNG50.

\subsection{Why do half of the TNG50 jellyfish have, or not have, cold gas today?}
\label{sec:results_end}

\begin{figure}
    \includegraphics[width=\columnwidth]{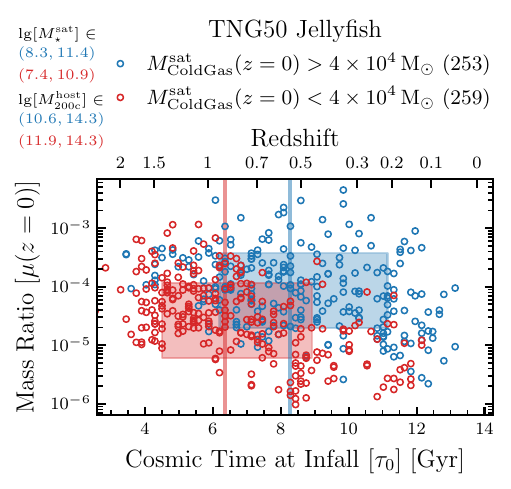}   
    \caption{
    {\bf Why do some jellyfish retain significant amounts of cold gas until $z=0$, while others do not?}
    The shaded regions and vertical lines mark the 16th/84th percentiles and medians of the distributions, respectively. The TNG50 jellyfish with cold gas masses below our resolution limit at $z=0$ ($\mcgassat <4\times10^{4}\, \msun$, red circles) typically have earlier infall times $\tau_0$ and lower mass ratios $\mu \equiv \mstarsat / \mvirhost(z=0)$ than the jellyfish with $\mcgassat > 4\times10^{4}\, \msun$ today (blue circles).
    }
    \label{fig:gas-nogas_mu_infalltime}
\end{figure}

The 512 jellyfish galaxies provided by TNG50 span orders of magnitude in their $z=0$ stellar mass, host mass, and importantly their (cold) gas mass (Fig.~\ref{fig:TNG50_selection_hists}). Why do half of the jellyfish galaxies retain significant amounts of cold gas until $z=0$, while the others do not? 

As a reminder, in this paper we analyse TNG50 satellite galaxies that survive, in terms of their galaxy stellar mass, through $z=0$ (see \S\ref{sec:meth_zoon} and \S\ref{sec:meth_sample} for more details). 
In Fig.~\ref{fig:gas-nogas_mu_infalltime}, we show the satellite-to-host mass ratio $\mu$ vs. the infall time for the population of jellyfish branches that end up with cold gas masses above (blue circles) or below (red circles) our resolution limit ($4\times10^{4}\, \msun$) at $z=0$. Here, the 16/84th percentiles and medians are marked with the shaded regions and vertical lines respectively. We note that the results remain qualitatively similar when using the satellite-to-host mass ratio at infall rather than at $z=0$.

The average infall occurs $\sim2$~Gyr earlier for those jellyfish galaxies with little to no cold gas remaining than for those that still retain some cold gas at $z=0$. While the host halo masses might have not had as much time to grow at earlier times and the $z=0$ mass ratios are $\mu\sim10^{-3} - 10^{-5}$, most of the early infallers (with infall times $\gtrsim5$~Gyr ago) have had enough time until $z=0$ to undergo secular and environmental processes to lose their cold gas. Even if these galaxies required multiple pericentric passages to lose their gas, they have had enough time before $z=0$ to have done so. Conversely, the largest majority of late-infalling jellyfish (i.e. with infall times as recent as a few Gyr ago) that have lost their gas by $z=0$ exhibit very low $z=0$ mass ratios (in the range $\mu \sim 10^{-4} - 10^{-6}$), whereas those with cold gas today typically have $\mu \sim 10^{-3}-10^{-5}$, either because they are more massive or because they orbit in less massive hosts. 

We speculate that the galaxies with non-vanishing cold gas masses that remain satellites (i.e. do not become backsplash galaxies) would eventually lose all their cold gas, i.e. if the simulation ran longer in time. 

Whereas the characterization of Fig.~\ref{fig:gas-nogas_mu_infalltime} is not surprising, it reminds us that, the longer a satellite has interacted with its host, the more time environmental processes, such as RPS, have had to act upon it. And even though some secondary effects may be in place -- such as galaxy selection, orbital trajectories, numbers of pericentric passages, edge-on vs. face-on orientation of the satellite as it falls into the host, and/or satellite-satellite interactions -- this zeroth-order picture is in line with what has already been quantified by \citet{Donnari2021,Joshi2021} for all TNG simulations: satellites that have spent more time in their hosts are more likely to be quenched compared to those that are still infalling or on their first infalling trajectory.

\subsection{For how long is ram pressure stripping in action?} \label{sec:results_long}

\begin{figure*}
    \includegraphics[width=\textwidth]{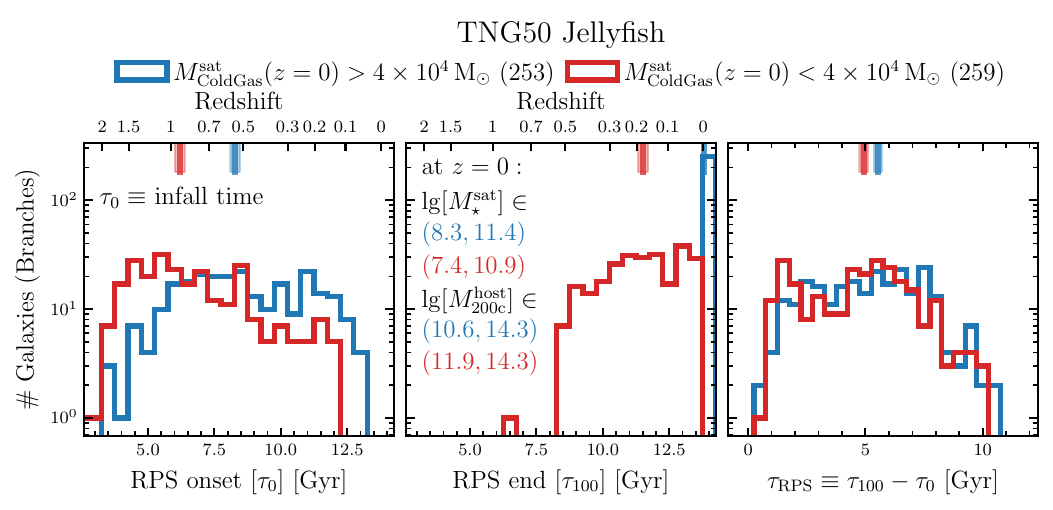}  
    \caption{
    {\bf Distributions of onset, end, and duration of ram-pressure stripping (RPS) for TNG50 jellyfish with cold gas masses at $z=0$ above (blue) and below (red) our resolution limit}. The onset of RPS $\tau_{0}$ (left panel) is defined as cosmic time at infall. The end of RPS $\tau_{100}$ (center panel) is defined as when the cold gas mass drops below our resolution limit ($\mcgassat \lesssim 4\times10^{4}\, \msun$) or the end of the simulation at $z=0$, if the galaxy always has $\mcgassat \gtrsim 4\times 10^{4}\, \msun$. The total RPS timescale $\tau_{\rm RPS}$, i.e. the total duration of RPS (right panel), is the difference between the end and onset of RPS. The jellyfish with substantial cold gas masses at $z=0$ all have $\tau_{100} = 13.8$~Gyr (the end of the simulation) by definition, causing the $\tau_{\rm RPS}$ distribution to be a reflection of the $\tau_0$ distribution. The medians and $1\sigma$ errors of the distributions are marked by the hash marks and shaded regions on the top $x$-axis. For the jellyfish with $\mcgassat(z=0) < 4\times10^{4}\, \msun$, the $\tau_{\rm RPS}$ distribution appears bimodal, with peaks at $\approx1.5-2.0$ and $4.5-6.5$~Gyr. We examine this distribution in detail in \S\ref{sec:results_long_phys}.
    }
    \label{fig:gas-nogas_jellyfish_tau_hists}
\end{figure*}

\begin{figure*}
    \includegraphics[width=\textwidth]{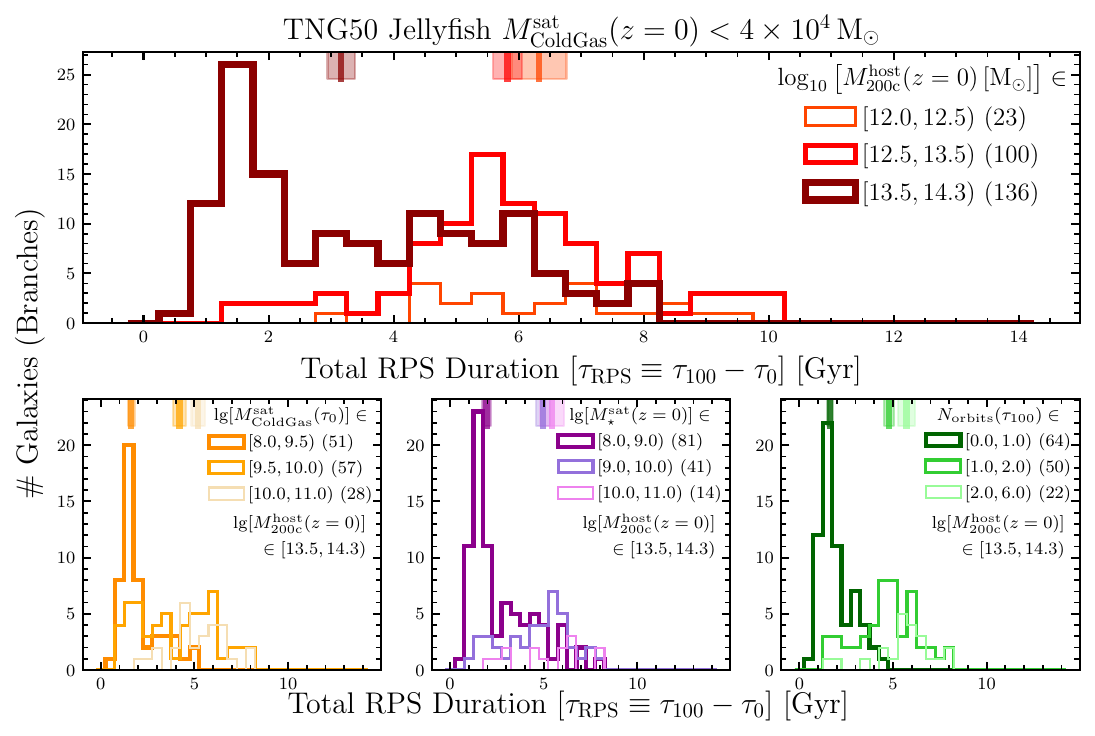}
    \caption{
    {\bf Given that a given jellyfish loses all of its cold gas, what determines how long it will take?}
    Of all the TNG50 jellyfish with $\mcgassat(z=0) < 4\times10^{4}\, \msun$, we bin the ram-pressure stripping (RPS) timescales $\tau_{\rm RPS}$ by host mass $\mvirhost(z=0)$ in the top panel. Here, only cluster mass hosts (dark red, $\mvirhost = 10^{13.5-14.3}\, \msun$, 8 hosts in total) have many jellyfish with both short and long RPS timescales. Then in the bottom panels we further bin the jellyfish orbiting in cluster-mass hosts by satellite cold gas mass at infall $\mcgassat(\tau_0$) (bottom left), satellite stellar mass today $\mstarsat(z=0)$ (bottom middle), and the number of orbits by the end of RPS $N_{\rm orbits}(\tau_{100})$, where the number of orbits is the number of apocentric passages, and the end of RPS ($\tau_{100}$) is the first time the satellite's cold gas mass falls below our resolution limit (see text for details). In all panels, the number of galaxies within each histogram is in parentheses in the panel legend; the medians and $1\sigma$errors are marked by the hash marks and shaded regions on the top $x$-axis.
    }
    \label{fig:tstrip_multipanel}
\end{figure*}

We are hence ready to quantify for how long RPS acts or has acted on TNG50 jellyfish galaxies. 

Fig.~\ref{fig:gas-nogas_jellyfish_tau_hists} shows the distributions of the onset of RPS ($\tau_0$ i.e. infall time; left), the end of RPS ($\tau_{100}$; middle), and the duration of RPS ($\tau_{\rm RPS}$; right), for TNG50 jellyfish with cold gas masses below (red) and above (blue) our resolution limit ($4 \times 10^{4}\, \msun$) at the current epoch. 

As we have seen in Fig.~\ref{fig:gas-nogas_mu_infalltime} and now again in the left panel, the jellyfish with little to no cold gas at $z=0$ are typically early infallers, with a majority falling in at $\tau_0 \approx 4.5-7$~Gyr after the Big Bang and a tail of late infallers at $\tau_0 \gtrsim 10$~Gyr after the Big Bang. After the RPS onset, the TNG50 jellyfish continue losing cold gas until sometime in the past few billion years ($\tau_{100} \approx 9-14$~Gyr). We only select galaxies which have been jellyfish since $z=0.5$, so we exclude galaxies that have been totally stripped of cold gas before $z = 0.5$. 

Finally, according to TNG50, the distribution of the RPS duration ($\tau_{\rm RPS}$, right panel) can be very wide, even for both subsets of jellyfish galaxies. For the jellyfish with substantial cold gas masses at $z=0$ all have $\tau_{100} = 13.8$~Gyr (the end of the simulation) by definition, causing the $\tau_{\rm RPS}$ distribution to be a reflection of the $\tau_0$ distribution. Thereby for these $z=0$ gaseous jellyfish, environmental effects, and hence RPS, have acted on them for as many as billions of years. For the jellyfish with cold gas masses below our resolution limit at $z=0$, the $\tau_{\rm RPS}$ distribution appears somewhat bimodal (see below for more details). Among these jellyfish, a fraction have undergone RPS for about $\approx 1.5-2.5$~Gyr and a larger fraction has undergone RPS for as long as $\sim4.5-7.5$~Gyr. As a reminder, these numbers represent the maximum time span over which RPS has acted (\S\ref{sec:meth_measurements_timings}); we see in the next Sections whether RPS may be more effective on shorter timescales.

\subsubsection{Physical origin of the diversity of RPS duration} \label{sec:results_long_phys}

What are the important factors in determining how long a given jellyfish needs to be stripped of its cold gas? We focus from now on only on those jellyfish that have cold gas masses below our resolution limit at the current epoch $\mcgassat(z=0) < 4 \times 10^{4}\, \msun$. 

In the top panel of Fig.~\ref{fig:tstrip_multipanel}, we extract the distribution of the duration of RPS $\tau_{\rm RPS}$ for the TNG50 jellyfish without substantial cold gas today, stacked by halo mass $\mvirhost$ of their current host. The number of jellyfish (not the number of hosts) belonging to each host mass bin is in parentheses in the upper right corner. 

Jellyfish in clusters ($\mvirhost = 10^{13.5-14.3}\, \msun$, dark red histogram) exhibit the shortest median RPS duration (vertical dark red line), although the distribution peaks at even shorter time spans: $\tau_{\rm RPS} \sim 1.5-2$~Gyr. Then there is a valley at intermediate stripping times $\tau_{\rm RPS}\sim2.5-4$~Gyr, followed by a slight increase from $\tau_{\rm RPS} \sim 4.5-6$~Gyr. The longest timescale for any jellyfish in this host mass bin is 8~Gyr. The jellyfish in groups ($\mvirhost = 10^{12.5-13.5}\, \msun$, red histogram) show RPS timescales that are single-peaked, with the median and mode coinciding at $\tau_{\rm RPS}\approx 5.5$~Gyr. While not shown but explicitly checked, jellyfish in group-mass hosts typically require at least 2 pericenteric passage to become fully stripped of cold gas. There are only $\approx 10$ ($\approx 10$~per~cent) galaxies in this host mass bin with stripping times shorter than 4~Gyr. This agrees with our earlier argument that RPS is more effective in higher host masses. Moreover, the jellyfish in approximately Milky-Way mass halos ($\mvirhost = 10^{11.5-12.5}\, \msun$, light red) require at least 4~Gyr, or in some cases much longer, to be fully stripped of their cold gas today. In general as RPS becomes more effective with increasing host mass, there are typically more jellyfish galaxies per host with increasing host mass \citep[see also fig.~14 from][]{Zinger2023}. However, even for these MW-mass hosts, the satellite-to-satellite variation is very large: there are TNG50 jellyfish that undergo RPS for as long as 10 billion years in both group- and MW-mass hosts, and as long as 8 billion years for cluster-mass hosts. 

The trend whereby shorter RPS time spans occur, on average, for satellites in more massive hosts is consistent with expectations described in \S~\ref{sec:intro}. However, here we quantify it for the first time with a large number of jellyfish across a wide range of host and satellite masses. Moreover, this trend is in place (physically vs. hierarchical growth of structure) even though more massive hosts exhibit in fact overall more recent infall times of their $z=0$ satellites than less massive hosts (not shown but explicitly checked). 

In the bottom row of Fig.~\ref{fig:tstrip_multipanel}, we focus on the TNG50 jellyfish with no remaining cold gas at $z=0$ in the 8 cluster hosts ($\mvirhost = 10^{13.5-14.3}\, \msun$) to investigate which additional physical properties imprint secondary trends on the duration of RPS. In practice, we show the $\tau_{\rm RPS}$ distributions binned by the satellites' cold gas mass at infall (left), stellar mass at $z=0$ (middle), and number of apocentric passages by $\tau_{100}$ (right). 

In the bottom left panel of Fig.~\ref{fig:tstrip_multipanel}, we see that satellites with the smallest (dark orange) and largest (light orange) amount of cold gas at infall are both single peaked at $\tau_{\rm RPS}\approx 1.5$ and $5$~Gyr respectively. Conversely, the intermediate bin (orange) has an approximately uniform distribution from $\tau_{\rm RPS} \approx1.5-6$~Gyr. Galaxies with less strippable material at infall tend to have shorter stripping durations. While not shown but explicitly checked, this trend remains for galaxies in a fixed host halo and galaxy stellar mass bin. 

In the bottom middle panel, lower mass jellyfish (dark purple) are typically stripped of all their cold gas faster, on average in $1-2$~Gyr, although a non-negligible fraction of them still require $3-6$~Gyr to be fully stripped of their cold gas. The intermediate (purple) and high (light purple) bins of satellite's stellar mass have similarly flat distributions with both a median RPS duration of $\sim 5$~Gyr. 

Lastly in the bottom right panel of Fig.~\ref{fig:tstrip_multipanel}, we show that satellites with the shortest RPS durations are those with the fewest orbits by being totally stripped $N_{\rm orbits}(\tau_{100})$, where $N_{\rm orbits}(\tau_{100})$ is the number of apocentric passages before $\tau_{100}$. The jellyfish with the shortest RPS duration are those that get stripped of all their cold gas before or immediately after their first pericentric passage (dark green), whereas satellites that require longer RPS time spans to be fully stripped of their cold gas are characterized by more than one apocentric passage (green and light green histograms).

\subsection{When and where does ram pressure strip most of a jellyfish's cold gas?} \label{sec:results_where}

\begin{figure*}
    \includegraphics[]{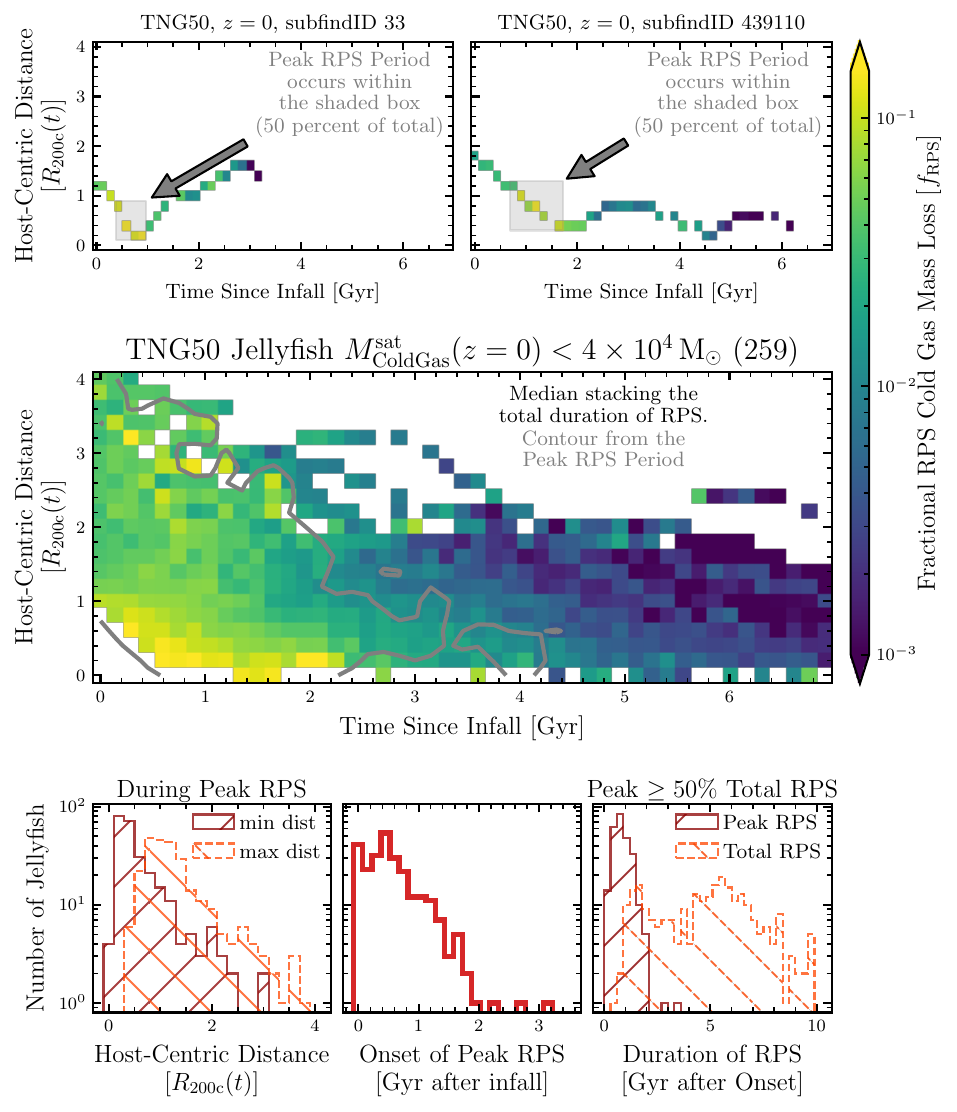}
    \caption{
    {\bf When after infall, where within the host halo, and for how long does 50~per~cent of RPS occur?}
    Top panels: the cold gas mass loss due to RPS at each snapshot normalized by the total amount of cold gas lost due to RPS through the satellite's life for two example galaxies (i.e. fractional RPS loss as per equation~\eqref{eqn:fRPS}). The peak RPS period -- the minimum amount of time for 50 percent of the total RPS to occur (equation~\eqref{eqn:peakRPS}) -- is within the gray box. The galaxy in the top left panel is the same as in Figs.~\ref{fig:SCGM_evolution},~\ref{fig:tracer_evolution}. 
    Central panel: Median stacking of the fractional RPS loss of all 259 TNG50 Jellyfish with $\mcgassat(z=0)<4\times10^{4}\, \msun$. The gray contour marks the phase space obtained when stacking only the peak RPS periods of all 259 jellyfish. 
    Bottom panels: Histograms detailing the peak RPS period of the 259 TNG50 Jellyfish with $\mcgassat(z=0) < 4\times10^{4}\, \msun$. Bottom left panel: the distributions of minimum (dark red, `/' hatch, solid outline) and maximum (light red, `\textbackslash' hatch, dashed outline) host-centric distances [$\rvir(t)$] within the peak RPS period. Bottom center panel: the distribution of onsets of the Peak RPS periods [Gyr after infall]. Note that the total RPS time span $\tau_{\rm RPS}$ begins at infall which is at 0 Gyr on this plot. Bottom right panel: the distributions of the peak RPS (dark red, `/' hatch, solid outline) and total RPS (light red, '\textbackslash' hatch, dashed outline) time spans. Note that the total RPS time span distribution is identical to that in the right panel of Fig.~\ref{fig:gas-nogas_jellyfish_tau_hists}.
    }
    \label{fig:fractionalRPSloss}
\end{figure*}

While most TNG50 jellyfish are stripped of their cold gas within $\sim1-7$~Gyr after infall (Fig.~\ref{fig:gas-nogas_jellyfish_tau_hists} right panel; Fig.~\ref{fig:tstrip_multipanel} top panel), the amount of gas being stripped is not constant throughout this period. The halo gas is denser at closer distances to the central galaxy, and jellyfish move faster at closer distances while they are deeper in their host's potential well. Both of these effects increase the ram pressure stripping acting on satellites at closer distances. We quantify this increase in RPS with decreasing distance in Fig.~\ref{fig:fractionalRPSloss}. In the top panels for two example galaxies, we show the fractional RPS loss $f_{\rm RPS}(t_i)$ -- the amount of cold gas lost due to RPS since the last snapshot $\Delta t_{{\rm snap}_i} = t_{i} - t_{i-1}$, normalized by the total amount of cold gas lost due to RPS in the satellite's life:
\begin{equation} \label{eqn:fRPS}
    f_{\rm RPS}(t_i) = \ddfrac{\int_{t_{i-1}}^{t_i} {\rm RPS}(t)\ {\rm d}t}{\int_{\tau_0}^{\tau_{100}} {\rm RPS}(t)\ {\rm d}t}
\end{equation}
where $\tau_0$ and $\tau_{100}$ define the total time span of satellite RPS. In general, the fractional RPS increases as a jellyfish approach pericenter, followed by a decrease as it approaches apocenter. For galaxy with SubhaloID 439110 (right panel), the fraction of total gas lost is higher during the first pericentric passage compared to the second because $\mcgassat$ -- the total amount of cold gas able to be stripped -- is an order of magnitude higher at infall than after its first orbit (at apocenter). There is still an increase in fractional RPS during the second pericentric passage compared to at apocenter, but the majority of RPS for this jellyfish occurs during the first infall through pericentric passage. 

We characterize the period of most effective RPS, the peak RPS period, by finding the minimum amount of time required for 50~per~cent of the total cold gas loss via RPS to occur. That is, we minimize the difference in bounds $(t_{\rm stop} - t_{\rm start})$ such that the integral of the fractional RPS $f_{\rm RPS}(t)$ is at least 50 percent:
\begin{equation} \label{eqn:peakRPS}
    \text{peak RPS} \coloneqq \text{MIN}\left(t_{\rm stop} - t_{\rm start}\right) \left| \int_{t_{\rm start}}^{t_{\rm stop}} f_{\rm RPS}(t)\ {\rm d}t \geq 0.5 \right. .
\end{equation}
We highlight the peak RPS periods for the two examples galaxies in Fig.~\ref{fig:fractionalRPSloss}, top, with gray boxes, which in both cases occur during the first infall towards pericenter. 

In the central panel of Fig.~\ref{fig:fractionalRPSloss}, we stack all 259 TNG50 jellyfish with $\mcgassat(z=0) < 4\times 10^{4}\, \msun$, taking the median fractional RPS loss in the bins with more than one galaxy. Additionally, the gray contour denotes the phase space region obtained when stacking only the peak RPS periods of the jellyfish. Based on the fractional RPS loss (color of bins) and the peak RPS contour, a majority of the RPS occurs within the first few Gyrs after infall and over a wide range of host-centric distances. At a fixed time since infall (single column), there is a higher fractional RPS loss at closer distances. However and especially at times $\lesssim 2$~Gyr after infall, there is a significant amount of RPS occurring at large host-centric distances, up to $\approx 3\rvir$. At later times $\gtrsim 2.5$~Gyr after infall, the peak RPS only occurs at closer host-centric distances $\approx 0.2-1.0\rvir$. The smallest host-centric bin $< 0.1\rvir$ is largely unpopulated (no color) or with few galaxies (not shown but checked). This means that their pericentric passages are at distances $\gtrsim 0.1\rvir$, and that they are being stripped of their cold gas in the halos rather than directly into central galaxy. This again supports the claim that tidal stripping is likely not a significant mechanism for cold gas removal for this jellyfish sample.

In the bottom panels of Fig.~\ref{fig:fractionalRPSloss}, we show the distributions of the peak RPS period quantities. In the bottom left panel, we show the minimum (dark red, '/' hatch, solid outline) and maximum (light red, '\textbackslash' hatch, dashed outline) host-centric distances during the peak RPS periods. The minimum peak RPS distance distribution has its peak (mode) at $0.3\rvir$, and the median (16, 84 percentiles) are $0.43\, (0.22,\ 1.1)\ \rvir$. The maximum peak RPS distance distribution peaks at $\rvir$, with median (16, 84 percentiles) at $1.2\, (0.75,\, 1.9)\ \rvir$. These distributions reflect that the peak RPS period starts at large distances in the halo (which has been discussed in, e.g., \citealp{Bahe2013,Zinger2018}) and continues until approximately the pericentric approach, which for our sample of jellyfish that lose all cold gas by $z=0$ tends to be at $\gtrsim 0.2\rvir$ \citep[see][for more details about TNG jellyfish at large distances $\dsathost>\rvir$]{Zinger2023}. The cold gas is being stripped in, and thereby deposited into, the host halos; we extensively expand on this in \S~\ref{sec:disc_CGM} and Fig.~\ref{fig:halos_totalRPS}.

In the bottom center panel, we see that the onset of the peak RPS occurs at or just after ($\lesssim 1$~Gyr) infall. Only $\approx15$~per~cent of these jellyfish galaxies begin their peak RPS period $>1$~Gyr after infall, suggesting that the infall time is a reasonable definition for the start of the total RPS time span.

In the bottom right panel, we show the peak RPS (dark red, '/' hatch, solid outline) and total RPS (light red, '\textbackslash' hatch, dashed outline) time spans. The two distributions here have different times of onset; the peak RPS onset is that given in the bottom center panel, while the total RPS onset is the infall time, which would be 0 in the bottom center panel. The total RPS time span is identical to that in Fig.~\ref{fig:gas-nogas_jellyfish_tau_hists}. While the total RPS duration spans a broad range of times $\approx1-7$~Gyr, the peak RPS period is much narrower, spanning only $\lesssim 2$~Gyr after onset. Thus while the total RPS time span may be quite long, a majority of the RPS occurs in a relatively short period. While not shown here, the distribution of peak to total RPS time span ranges from $\approx0.1-0.4$, with the mode and median at $\approx0.15$ and $0.20$ respectively. 

We note an alternative method for characterizing the effectiveness or peak RPS as the specific RPS (sRPS): RPS / $\mcgassat$. Typically for the TNG50 jellyfish without cold gas at $z=0$, the specific RPS + outflows is approximately constant before infall. At infall, the sRPS typically increases through pericenter and near apocenter either plateaus or decreases, sometimes to its pre-infall value. For the galaxies that lose all their cold gas only at or after second pericenter (subfindID 439110 in the top right panel of Fig.~\ref{fig:fractionalRPSloss} for example), the sRPS increases again and always reaches its maximum value at or shortly before $\tau_{100}$. See Appendix~\ref{app:tau} for more details and Fig.~\ref{fig:subfindID439110_MCGas_tracer_evolution} for an example.

\section{Discussion} \label{sec:disc}

\subsection{How do these jellyfish-based results generalize to all satellite galaxies?} \label{sec:disc_compare_all_satellites}

Throughout this paper, we have focused on jellyfish galaxies, as these are satellites with manifest signs of ongoing RPS. In particular, we have followed satellites along their evolutionary tracks through cosmic epochs and dubbed the inspected galaxies as jellyfish only if they have at least one jellyfish classification since $z=0.5$. This is when the temporal sampling of the images on Zooniverse transitioned from every $\approx1$~Gyr to every $\approx150$~Myr. We also restate that the images posted to Zooniverse used a fixed gas column density colorbar in the range $\Sigma_{\rm gas} \in 10^{5-8}\, \msun\, {\rm kpc}^{-2}$ and did not include background subtraction, mimicking a surface brightness limited sample. Hence to have been classified as a jellyfish galaxy, the stripped tails must have been dense enough to have been distinguishable against the background. Lastly, we expect that at any given snapshot, we miss $\approx30-40$~per~cent of jellyfish galaxies due to projection effects \citep{Yun2019,Zinger2023}.

As shown in Fig.~\ref{fig:TNG50_selection_hists} and discussed in \S~\ref{sec:results_branches}, jellyfish galaxies tend towards lower stellar masses $\mstarsat$, higher host masses $\mvirhost$, and lower satellite-to-host mass ratios $\mu$ at $z=0$ compared to the inspected galaxies and the general $z=0$ population of satellites. Conversely, the inspected galaxies that were \textit{not} identified as jellyfish tend towards the opposite. First 163 of the 1,031 ($\approx15$~per~cent) non-jellyfish galaxies have stellar masses $\mstarsat > 10^{10}\, \msun$, and may have experienced phases of kinetic AGN feedback, ejecting much of their gas. Of the lower mass non-jellyfish galaxies $\mstarsat < 10^{10}\, \msun$, we affirm that many of these satellite galaxies are still undergoing or have undergone RPS (see Figure~\ref{fig:halos_RPS}), and the question becomes why have they not been identified as jellyfish. These non-jellyfish galaxies have a median satellite-to-host mass ratio $\mu = 7.8\times10^{-4}$, $\approx 15$~times higher than that for the jellyfish $\mu = 5.0\times10^{-5}$. We generally expect and have shown in Fig.~\ref{fig:tstrip_multipanel} that with increasing satellite-to-host mass ratio RPS is weaker and acts over longer time spans. Accordingly, some of the gaseous tails may not have been identifiable in gas column density compared to the ambient medium. Moreover, the non-jellyfish galaxies have a median infall time at $9.2\pm1.2$~Gyr, $\approx1.7$~Gyr later than the jellyfish galaxies at $7.5\pm1.0$~Gyr. So it is also possible that these late-infalling non-jellyfish have not yet had enough time to undergo enough RPS to form the recognizable tails, although the timescales associated with the appearance and disappearance of the jellyfish tails is largely unconstrained \citep{Smith2022}. 

While 259/512 ($\approx51$~per~cent) of the jellyfish galaxies have cold gas masses below our resolution limit at $z=0$, this is only the case for 125/1,031 ($\approx12$~per~cent) of the non-jellyfish galaxies. Then how can these 125 gas-less satellites have lost all of their cold gas without being identified as jellyfish? Of the galaxies with $\mcgassat(z=0) < 4\times 10^{4}\, \msun$, the RPS duration for jellyfish includes both short and long time spans $\tau_{\rm RPS} \approx 1.5-8$~Gyr, while for the non-jellyfish the time spans are only long $\tau_{\rm RPS} \approx 3.5-7.5$~Gyr. Again, this demonstrates that the RPS for the non-jellyfish with higher mass ratios is slower, potentially causing the gaseous tails to be unidentifiable. Furthermore, these $z=0$ gas-less non-jellyfish tend to have even earlier infall times than their jellyfish counterparts. In fact, almost all of the non-jellyfish have infall times before $z=0.5$, before the temporal sampling of the images on Zooniverse transitioned from every $\approx1$~Gyr to every $\approx150$~Myr. So before $z = 0.5$, we may be missing some jellyfish simply by not inspecting their images often enough. Based on the statistical, physical differences between the general jellyfish and non-jellyfish galaxies, and that we may be missing some high-redshift, jellyfish-like galaxies, we conclude that our primary sample of jellyfish galaxies is pure, but perhaps not complete. And when generalizing the results of the RPS time spans from the jellyfish to all satellites, the time spans would only increase. We have also checked that the peak RPS periods are slightly longer and still occur in the halos for the non-jellyfish galaxies. However, our results only apply to first-infalling galaxies, i.e not to pre-processed galaxies, which is more significant for less massive satellites in more massive hosts. Extending this analysis to pre-processed galaxies would require distinguishing how much RPS occurs in each host, and when the infalling group's intragroup medium gets stripped.

\begin{figure*}
    \includegraphics[width=0.4\textwidth]{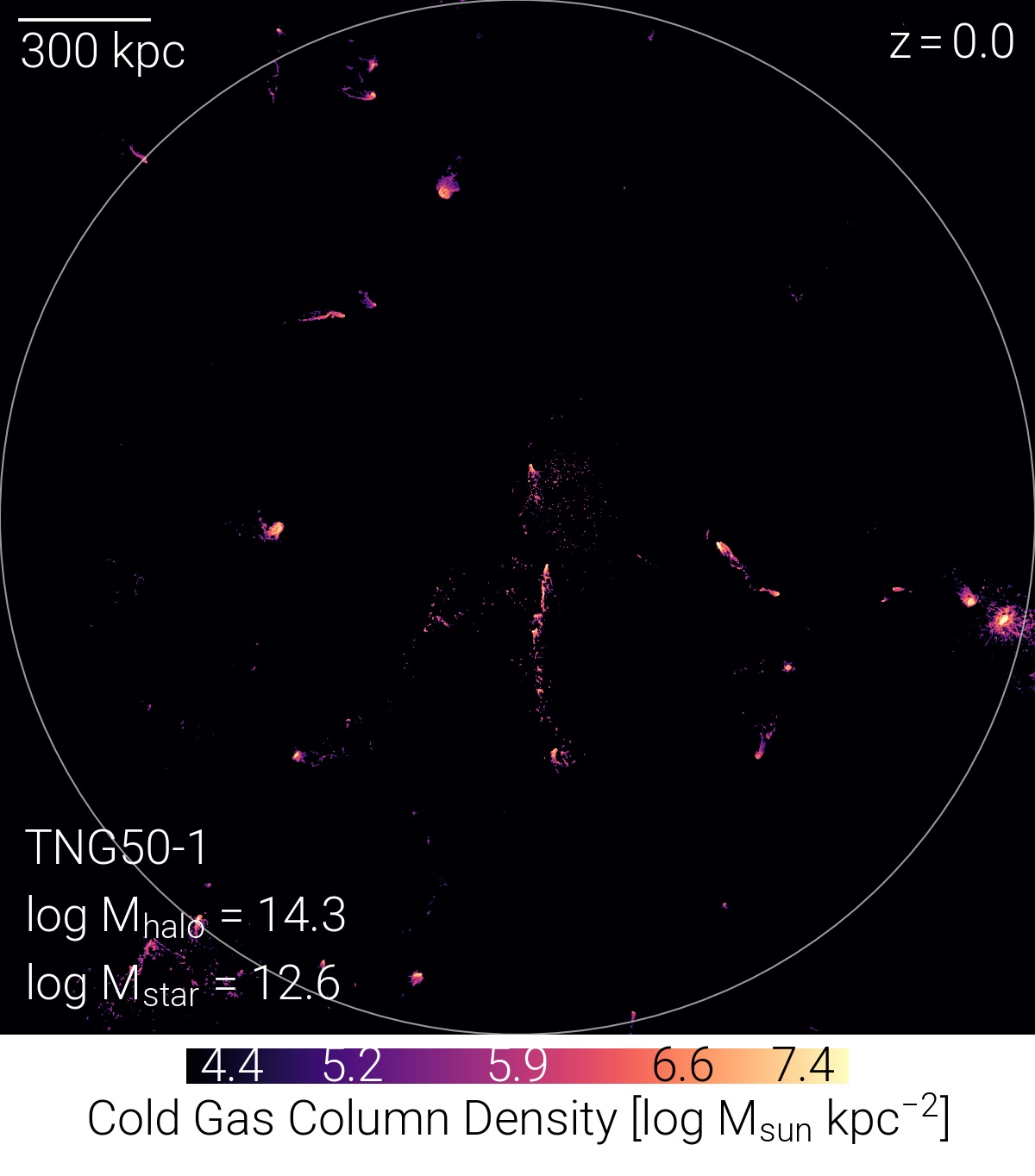}
    \includegraphics[width=0.4\textwidth]{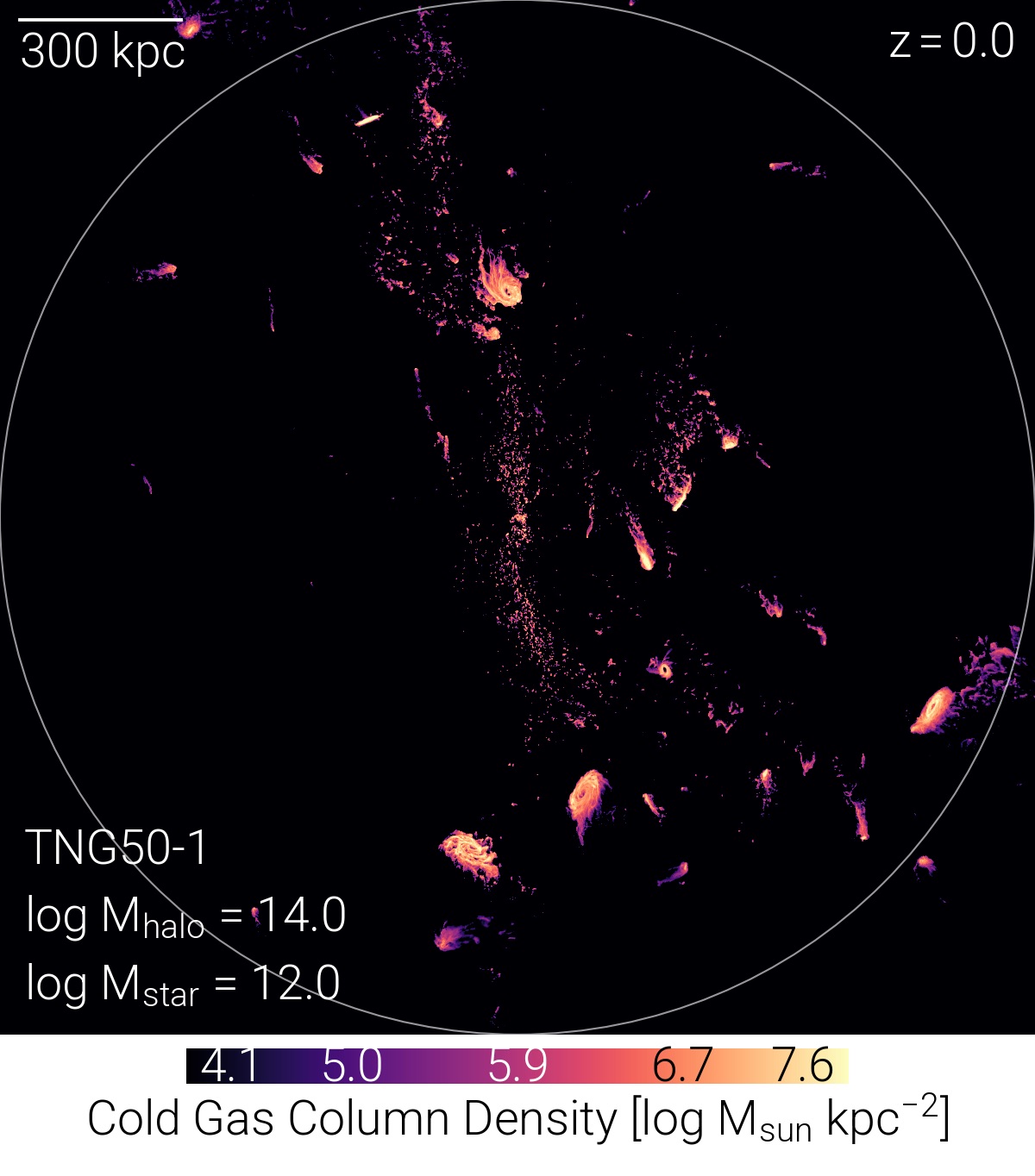}
    \includegraphics[scale=1.0]{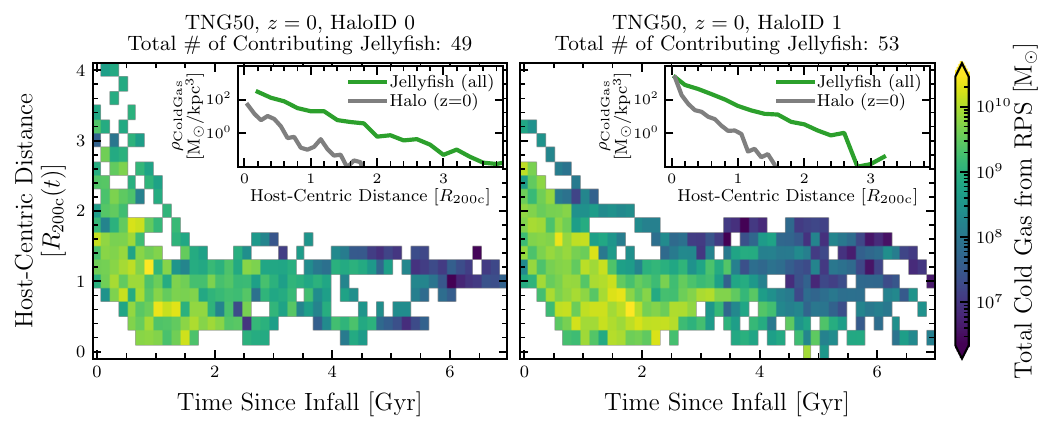}
    \caption{
    {\bf TNG jellyfish deposit a significant amount of cold gas into their host halos.}
    For the most (left) and second most (right) massive clusters in TNG50 ($\mvirhost(z=0) \approx 10^{14}\, \msun$), we plot the cold gas column density at $z=0$ (top panels; all halo gas with temperatures $\leq 10^{4.5}$~K, including gas gravitationally bound to satellites) and the total amount of cold gas deposited in the host halos from all TNG50 jellyfish in bins of host-centric distance and time since infall (bottom panels). For both halos, the total amount of cold gas deposited into the halos from ram-pressure stripped jellyfish is $\sim 10^{12}\, \msun$ over the last 5 billion years. In the insets (bottom panels), we compare the radial distributions of the cold gas deposited via RPS from all jellyfish at $z\lesssim0.5$ (green) with the cold gas that exists in and around the halos at $z=0$ (gray, excluding cold gas bound to satellite galaxies). Together with Fig.~\ref{fig:halos_RPS}, this shows that jellyfish, and more generally satellites, contribute a significant amount of cold gas into their host halos.
    }
    \label{fig:halos_totalRPS}
\end{figure*}

\subsection{The connection between ram pressure stripping and quenching timescales} \label{sec:disc_quench}

The 259 TNG50 jellyfish that are gas-poor at $z=0$ were star forming galaxies -- on the star forming main sequence (SFMS) -- before infall and are instead quenched at $z=0$ -- at least than one dex below the SFMS. The question becomes, when between the RPS onset at infall and its end at $\tau_{100}$ do these jellyfish quench. We calculate the amount $f_{\rm RPS}(<\tau_{\rm quench})$ of RPS that has already occurred by the time of last quenching $\tau_{\rm quench}$ via
\begin{equation} \label{eqn:tau_quench}
    f_{\rm RPS}(<\tau_{\rm quench}) = \int_{\tau_0}^{\tau_{\rm quench}}f_{\rm RPS}(t)\, {\rm d}t 
\end{equation}
where the fractional RPS $f_{\rm RPS}(t)$ is defined in equation~\eqref{eqn:fRPS}, and $\tau_{\rm quench}$ is the last time that the galaxy falls at least one dex below the SFMS for the last time \citep{Pillepich2019,Joshi2021,Donnari2021}. 

On average, the jellyfish do not quench until $\gtrsim 99$~per~cent of the total RPS has occurred. Only 5/259 ($\approx2$~per~cent) of the jellyfish quench before $f_{\rm RPS}(<\tau_{\rm quench})=97$~per~cent. Moreover, these jellyfish have already lost $\gtrsim 98$~per~cent of their cold gas by the time they quench. Of the 259 jellyfish galaxies with $\mcgassat(z=0) > 4\times10^{4}\, \msun$), only 74 ($\approx 30$~per~cent) have quenched whereas the others are still forming stars \citep[see also][]{Goeller2023}. These quenched jellyfish also have already lost $\approx 98$~per~cent of their cold gas before quenching. While the peak period of RPS typically occurs during the first infall through pericenter, lasting $\lesssim 2$~Gyr, the jellyfish do not quench for the last time until nearly all of their cold gas has been stripped on time spans that can be $\gtrsim 5$~Gyr after infall. This does not necessarily imply that the galaxies are on the SFMS for the entire duration between infall and $\tau_{100}$, but instead that they quench for the last time only after being stripped of almost all of their cold gas. Jellyfish galaxies are able to continue forming stars well after infall and after they have lost almost all of their cold gas due mostly to RPS.

To define a quenching timescale, one also needs to define the onset of quenching \citep[See][for a review of various definitions used in the literature]{Cortese2021}. If we assume the infall time $\tau_0$ as the onset of quenching, then the distribution of quenching timescales is approximately the same as the RPS timescale distribution in Fig.~\ref{fig:gas-nogas_jellyfish_tau_hists} (right panel, red histogram). Thus, the quenching timescales for the TNG50 jellyfish studied here and without cold gas at $z=0$ range from $\approx 1-7$~Gyr after infall. However, we note that many of these jellyfish undergo brief ($\lesssim 1$~Gyr) bursts of star-formation between infall and first pericentric passage \citep{Goeller2023}. While it may seem counter-intuitive for the onset of quenching -- in this case, infall -- to be directly before a burst of star-formation, this starburst coincides exactly with the time span that most jellyfish incur their peak gas loss due to RPS (Fig.~\ref{fig:fractionalRPSloss}). Thus the RPS and burst of star-formation may act together and enhance each other to remove cold gas from jellyfish, eventually quenching them. This is consistent with the satellite post-starburst quenching scenario, where ram pressure induces a burst of star-formation before the satellite eventually quenches and has signature of a post-starburst galaxy \citet{Poggianti2017,Gullieuszik2017,Vulcani2020,Grishin2021,Werle2022}.

\subsection{RPS deposits satellite ISM into the halo} \label{sec:disc_CGM}

\begin{figure}
    \includegraphics[width=\columnwidth]{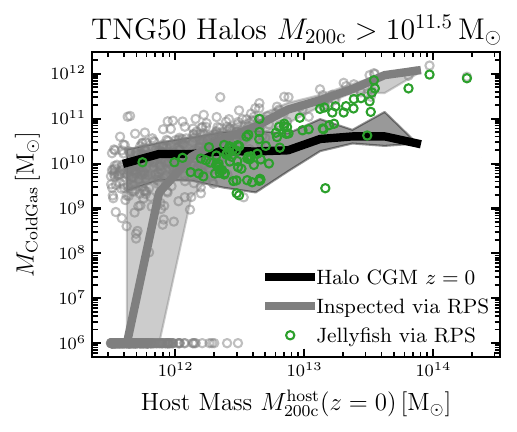}
    \caption{
    {\bf TNG satellites (both jellyfish and inspected satellite branches) are a significant source of cold gas for their host halos.}
    We extend the analysis from Fig.~\ref{fig:halos_totalRPS} now to include all inspected branches and hosts down to mass $\mvirhost = 10^{11.5}\, \msun$. We compare the the cold circumgalactic medium (CGM) excluding cold gas bound to satellites (black curve and shaded region denote the median and 16/84th percentiles) to the amount of cold gas brought in over the past $\sim 5$~Gyr by jellyfish (green circles) and all inspected galaxies (jellyfish + non-jellyfish; gray circles, where the gray curve and shaded region denote the median and 16/84th percentiles) from ram pressure stripping (RPS). We place by hand the halos without any cold gas deposited from inspected galaxies at $M_{\rm ColdGas} = 10^{6}\, \msun$. At host masses $\gtrsim 10^{13}\, \msun$, the inspected satellite galaxies have brought more cold gas into the halos over the past $\sim 5$~billion years than exists in the CGM today.
    }
    \label{fig:halos_RPS}
\end{figure}

In addition to being stripped of their cold gas, jellyfish galaxies, and more generally satellite galaxies, also provide a source of cold gas to the halo. For a given host, how much cold gas comes directly from ram-pressure stripped jellyfish galaxies? 

In Fig.~\ref{fig:halos_totalRPS}, we combine all jellyfish galaxies (regardless if they have substantial cold gas at $z=0$ or not) in the the two most massive clusters $\mvirhost(z=0) \approx 10^{14}\, \msun$ in TNG50, and show the total amount of deposited cold gas in the time since infall -- host-centric distance space, similarly to Fig.~\ref{fig:fractionalRPSloss}. We also show the cold gas column density maps for these clusters for reference. The two clusters have hosted 49 and 53 total contributing jellyfish since approximately $z\sim0.5$, depositing a total of $\sim 10^{12}\, \msun$ of cold gas mass into the hosts. This is a substantial amount of cold gas: it is about one tenth of the total amount of gas in these halos at any given time, it is of a similar order of magnitude as the stellar mass of the central galaxy of the host at $z=0$, and it is orders of magnitude more than the amount of ionized and molecular gas that have been recently observed around several brightest central galaxies \citep[e.g.][]{McNamara2014,Russell2019}.

We can integrate the contributed cold gas along the host-centric distance, yielding the 1D distribution of the deposited cold gas from RPS in time since infall. For the most massive halo in TNG50, the median (16/84 percentiles) of the cold gas from RPS distribution occurs $0.9\, (0.3/1.8)$~Gyr after infall; for the second most massive halo, $1.2\, (0.4/2.4)$~Gyr after infall. Again, while some jellyfish contribute cold gas to the hosts over long periods $\gtrsim 3$~Gyr, a majority of the RPS occurs shortly after infall, qualitatively agreeing with the results discussed in \S~\ref{sec:results_where} shown in Fig.~\ref{fig:fractionalRPSloss} (bottom center and right panels). 

Additionally, we can integrate the contributed cold gas along the time since infall, yielding the host-centric radial distribution of deposited gas from RPS. For the most massive halo in TNG50, the median (16/84 percentiles) of the cold gas from RPS distribution occurs at $1.0\, (0.6/2.4)\, \rvir$; for the second most massive halo, $0.8\, (0.2/1.8)\, \rvir$. Thus, we see that the majority of contributed cold gas from RPS is deposited into the outskirts of the gaseous halos (i.e., CGM or ICM) of these most massive hosts in TNG50. In the figure insets, we compare the radial-density distributions of cold gas deposited via RPS from all jellyfish (green) with the cold gas that exists in the halos at $z=0$ (gray, excluding cold gas bound to satellites). For these two cluster-mass hosts, more cold gas has been brought into their halos via RPS over the last many billion years than exists in their intra-cluster media today. 

We extend this analysis in Fig.~\ref{fig:halos_RPS} now to include all inspected satellite branches (gray circles) and group- ($\mvirhost(z=0) \sim 10^{13}\, \msun$) and Milky-Way-mass ($\mvirhost(z=0)\sim 10^{12}\, \msun$) hosts. According to our analysis and to TNG50, over the past $\sim 5$ billion years satellite galaxies have deposited more than $10^{10}\,\msun$ of cold gas mass via RPS in the CGM of halos more massive than $10^{12.5\,\msun}$. The amount of cold gas in the CGM at $z=0$ (black circles) increases with halo mass until $\sim 10^{13.5}\, \msun$, and afterwards is approximately constant. The amount of cold gas deposited by inspected galaxies in low mass hosts $\mvirhost \lesssim 10^{12}\, \msun$ is bimodal, where many hosts have 0 inspected branches. Of the low-mass hosts with inspected branches, the amount of cold gas deposited by ram pressure stripping (RPS) increases with halo mass, which continues with all studied halo masses. 

Of the amount of cold gas deposited by RPS of the inspected galaxies, the relative contribution of jellyfish galaxies increases with halo mass, reflecting the trend that a higher percentage of inspected galaxies are jellyfish at the higher host masses (see Fig.~\ref{fig:TNG50_selection_hists}). At host masses $\gtrsim 10^{13}\, \msun$, the inspected galaxies have brought more cold gas into the halos over the past $\sim 5$~billion years than exists in the CGM today.

Thereby, we claim that jellyfish, and the more generally inspected or satellite galaxies, bring a significant amount of cold gas in the CGM/ICM of massive halos. The question then becomes, what happens to the stripped cold gas between being deposited and $z=0$. We speculate that this gas could either (i) remain cold in the CGM, (ii) remain cold and rain down on the central galaxy, (iii) mix and heat up with the surrounding hot medium, and/or (iv) be heated up and/or pushed outside of the halo by kinetic AGN feedback \citep[e.g.,][]{Ayromlou2023}. Conversely, one could start with the cold CGM clouds at $z=0$ and follow their histories back in time, quantifying how much came from satellites \citep{Nelson2020}. We postpone the task of quantifying the fate of the cold gas brought by satellites into the CGM around galaxies for a future work.

\section{Summary and Conclusions} \label{sec:sum}

In this work, we use the high-resolution, $\sim50$~Mpc magneto-hydrodynamical simulation TNG50 from the IllustrisTNG project to study the satellite-host interaction in a cosmological context for $\approx 5$ orders of magnitude in satellite stellar and host total mass. In particular, we quantify when, where, and for how long the ram pressure stripping (RPS) of cold gas occur, by focusing on jellyfish galaxies, i.e. satellites with manifest signs of RPS.

We use the results from \citet{Zinger2023}, which is a follow up from the pilot Zooniverse study from \citet{Yun2019}, to identify jellyfish galaxies via visual inspection. Namely, \citet{Zinger2023} report and discuss the visual inspection via Zooniverse of 53,610 satellite galaxies from TNG50 with $f_{\rm gas} < 0.01$ and $\mstarsat > 10^{8.3}\, \msun$, in the TNG50 simulation. For this paper, we track the 53,610 inspected images across cosmic time, finding a total of 5,023 unique galaxy branches. In the main analysis of this work, we focus on the galaxy branches that survive until $z=0$, were inspected in the Zooniverse project since $z \leq 0.5$, are satellites at $z=0$, have not been pre-processed, and have well-defined infall times; this returns a pure sample of 1,543 galaxies. 512 of these 1,543 branches ($\approx 33$~per~cent) are jellyfish galaxies, meaning that they were classified as a jellyfish galaxy for at least one snapshot since $z \leq 0.5$.

Compared to the inspected galaxies and general $z=0$ satellites with $\mstarsat > 10^{8.3}\, \msun$, the TNG50 jellyfish galaxies tend to have lower stellar masses, higher host masses, lower satellite-to-host mass ratios, and less gas (Fig.~\ref{fig:TNG50_selection_hists}). The tails of the jellyfish galaxies are made up of mostly cold gas $(\leq 10^{4.5}$~K) with similar metallicities to the gas within the stellar body, suggesting that the tails stem from the interstellar media (Figs.~\ref{fig:TNG50_temperature_poster},~\ref{fig:TNG50_metallicity_poster}), though the jellyfish tails may also be observable in, e.g., soft x-rays. 

We employ the Monte-Carlo-Lagrangian tracer particles to quantify the relative importance of each cold gas sink, namely SMBH accretion, star-formation (SF), gas heating, and ram-pressure stripping (RPS) + outflows. As individual galaxy tracks suggest, we assume that before infall, i.e. the first time the galaxy becomes a member of its Friends-of-Friends (FoF) host group, RPS+outflows category is dominated by outflows, and after infall RPS is dominant. Then we define the onset $\tau_0$ of RPS as the infall time and the end $\tau_{100}$ of RPS as either when the galaxy's cold gas mass falls below our resolution limit of $m_{\rm gas} \approx 4\times 10^{4}\, \msun$ ($f_{\rm gas} \lesssim 5\times10^{-4}$) or at the end of the simulation at $z=0$; then the total RPS time span $\tau_{\rm RPS}$ is the difference between $\tau_{100} - \tau_0$. With this sample of 512 jellyfish and method to measure RPS, our main results are:

\begin{itemize}
    \item For an individual example, we show that a single jellyfish branch loses all of its cold gas between infall and apocenter (Fig.~\ref{fig:SCGM_evolution}), and during this period RPS is the dominant channel of cold gas loss (Fig.~\ref{fig:tracer_evolution}). We check and find that RPS dominates the post-infall cold gas loss for all other jellyfish in the sample.
    \item Approximately half 259/512) of the jellyfish have been stripped of all cold gas by $z=0$. The jellyfish without cold gas at $z=0$ (i.e. with cold gas mass $<4\times10^4\,\msun$) tend to have smaller satellite-to-host mass ratios and earlier infall times than the jellyfish that retain some cold gas at $z=0$ (Figs.~\ref{fig:gas-nogas_mu_infalltime},~\ref{fig:gas-nogas_jellyfish_tau_hists}). 
    \item For the 259 jellyfish galaxies without cold gas at $z=0$, the total RPS durations span $\tau_{\rm RPS}\approx 1-7$~Gyr (Fig.~\ref{fig:gas-nogas_jellyfish_tau_hists}). The dominant factor for determining the RPS time span is the host mass, whereby jellyfish in higher-mass hosts have shorter RPS durations (Fig.~\ref{fig:tstrip_multipanel}, top panel). Secondarily, RPS durations decrease with satellite cold gas mass at infall, the stellar mass at $z=0$, and the number of orbits by $\tau_{100}$ (Fig.~\ref{fig:tstrip_multipanel}, bottom panels respectively).
    \item While the total RPS duration may be quite long, most jellyfish incur a majority of their cold gas mass loss via RPS within a short peak RPS period, beginning $\lesssim 1$~Gyr after infall and lasting $\lesssim2$~Gyr (Fig.~\ref{fig:fractionalRPSloss} top, bottom center, and bottom right panels). Typically this peak RPS period occurs within $\approx0.2-2\rvir$ of the host and during the first infall. 
    \item Jellyfish galaxies continue forming stars for billions of years after infall, until they have lost $\approx98$~per~cent of their cold gas mass. They quench for the last time only after $\approx99$~per~cent of the RPS has occurred (\S~\ref{sec:disc_quench}).
    \item In the two most massive $\sim 10^{14}\, \msun$ halos in TNG50, jellyfish galaxies contribute $\approx 10^{12}\, \msun$ of cold gas into the intra-cluster medium over the past $\sim 5$ billion years (ICM; Fig.~\ref{fig:halos_totalRPS}). The radial distribution of cold gas brought in via jellyfish RPS is significantly higher than the amount of cold gas existing in the ICM today. In fact, satellite galaxies deposit over the past $\sim 5$ billion years $\gtrsim 10^{10}\,\msun$ of cold gas in the CGM of $\gtrsim10^{12.5}\,\msun$ TNG50 halos (Figs.~\ref{fig:halos_RPS}). For massive hosts, this cold gas contribution is of the same order of magnitude as the stellar mass in the central galaxy today. Therefore, jellyfish galaxies, and the more general population of satellites, bring a significant amount of cold gas into the CGM/ICM of massive hosts.
\end{itemize}

In summary, we have shown that, according to TNG50, RPS is the dominant cause of loss of cold gas in satellites after they start to interact with their $z=0$ hosts and that satellite galaxies are significant contributors of cold gas to the CGM and ICM. RPS acts on infalling galaxies for very long periods of time, i.e. {\it many} billion years on average, even though the majority of the cold gas mass loss occurs faster, with half of the cold gas of satellites being stripped in the span of about 2 billion years or less. This cold gas is typically deposited by the satellites all the way from intermediate host-centric distances to beyond the virial radii of their hosts. 

We note that these results apply only to the satellite stellar and total host masses studied in this work, within the TNG model of galaxy formation. For the most massive satellites, $\mstar \sim 10^{10-11}\, \msun$, it is possible that their stellar potential is deep enough to retain some of their own CGM post-infall, shielding some of their ISM gas. At these masses, the TNG kinetic mode of SMBH feedback also becomes important, and is thought to dominate, along with RPS, the quenching of these satellites \citep{Donnari2020}. In a future work, we extend these results to more massive satellite and host masses using the upcoming TNG-Cluster project that focuses on massive hosts $\mvirhost \sim 10^{14-15.4}\, \msun$ using the TNG galaxy formation model.

\section*{Acknowledgements}
We thank the anonymous referee for the helpful suggestions that improved the quality and clarity of this manuscript.

ER is a fellow of the International Max Planck Research School for Astronomy and Cosmic Physics at the University of Heidelberg (IMPRS-HD). ER would like to acknowledge the following friends and colleagues (in alphabetical order) for useful comments and discussions that improved the quality of the manuscript: Luca Cortese, Morgan Fouesneau, Junia G{\"oller}, Max H{\"a}berle, Iva Momcheva, Nhut Truong, Hans-Walter Rix, Nico Winkel.

DN and MA acknowledge funding from the Deutsche Forschungsgemeinschaft (DFG) through an Emmy Noether Research Group (grant number NE 2441/1-1).

GJ acknowledges funding from the European Union’s Horizon 2020 research and innovation programme under grant agreement No. 818085 GMGalaxies.

The TNG50 simulation was run with compute time granted by the Gauss
Centre for Supercomputing (GCS) under Large-Scale Projects GCS-DWAR on the
GCS share of the supercomputer Hazel Hen at the High Performance Computing
Center Stuttgart (HLRS).

This publication uses data generated via the \href{https://www.zooniverse.org/}{Zooniverse.org} platform, development of which is funded by generous support, including a Global Impact Award from Google, and by a grant from the Alfred P. Sloan Foundation. We wish to extend our thanks to the team at Zooniverse with their advice and assistance in building and running this project. We also thank the thousands of volunteers who invested their time and effort to assist us in this project.

\section*{Data Availability and Software Used}
All TNG simulations are publicly available at \url{https://www.tng-project.org/} and described in \citet{Nelson2019}. All data from the Cosmological Jellyfish project are publicly available at \url{https://www.tng-project.org/data/docs/specifications/\#sec5\_3} and described in \citet{Zinger2023} and in this work. All codes used to analyze the TNG and Cosmological Jellyfish data and to produce the figures in this paper are publicly available at \url{https://github.com/ecrohr/TNG\_RPS}. 

Software used: {\sc Python} \citep{VanDerWalt2011}; {\sc IPython} \citep{Perez2007}; {\sc Numpy} \citep{VanDerWalt2011,Harris2020}; {\sc Scipy} \citep{Virtanen2020}; {\sc Matplotlib} \citep{Hunter2007}; {\sc Jupyter} \citep{Kluyver2016}.

This work made extensive use of the NASA Astrophysics Data System and \url{arXiv.org} preprint server.

\bibliographystyle{mnras}
\bibliography{references}

\appendix

\section{Tracking individual galaxies across epochs} \label{app:tracking}

In the Cosmological Jellyfish project, whose results we use here and are summarized and discussed by \citet{Zinger2023}, gas-map images were posted of TNG50 (and TNG100, though not discussed here) satellite galaxies meeting the criteria summarized in \S~\ref{sec:meth_zoon} at all 33 snapshots since $z=0.5$ (snapshots 99-67), plus at four additional snapshots corresponding to redshifts $z=0.7, 1.0, 1.5, 2.0$ (snapshots 59, 50, 40, 33). Many galaxies were inspected at multiple snapshots. However, for this work we focus on the unique evolutionary tracks, or branches, of individual galaxies using \sublinkgal \citep{Rodriguez-Gomez2015}. In practice, we load the main progenitor branch (MPB) of every galaxy inspected at $z=0$ (snapshot 99), saving the galaxies' subfindIDs at all previous snapshots. Then at each earlier inspection snapshot (98, 98, ..., 67, 59, 50, 40, 33), we check which galaxies' MPB have already been saved. If not, then we save the MPB and continue to the next snapshot. Within the 53610 inspected galaxy images in TNG50, there 5023 are unique branches. 

As summarized in \S~\ref{sec:meth_sample}, throughout this work we exclude galaxies if they do not exist at $z=0$, are backsplash galaxies at $z=0$, or have been pre-processed. For each branch that is not inspected at $z=0$ (snapshot 99), we load the main descendant branch (MDB). With the MDB, we find the last snapshot at which the galaxy exists, typically either when the galaxy merges with another more massive galaxy (subhalo coalescence) or at $z=0$. We are interested only in galaxies that exist as satellites at $z=0$, so we exclude 2,341 branches that do not exist as satellites at $z=0$. To determine whether the remaining galaxies that are satellites at $z=0$ are have been pre-processed, we examine both the galaxies' and their $z=0$ hosts' MPBs (technically the MPBs of their $z=0$ hosts' main subhalos). Then at each snapshot, we classify each galaxy's Friends-of-Friends (FoF) group membership into exactly one of three categories:
\begin{itemize}
    \item the main (central) subhalo of the group;
    \item a satellite of its $z=0$ host;
    \item a satellite of a group other its $z=0$ host.
\end{itemize} 
Then using these categorizations across the snapshots, we can determine whether the galaxy was pre-processed; if the galaxy spent at least $N_{\rm snaps} = 5$ consecutive snapshots as a satellite in a host -- other than its $z=0$ host -- of mass $\mvirhost > M_{\rm LowLim} = 10^{11}\, \msun$. If the galaxy instead spent these $N_{\rm snaps}$ snapshots in its $z=0$ host, then spent some snapshots as a central galaxy, before eventually being a satellite in the same group, then we do not consider the galaxy to be pre-processed and include the galaxy in the analysis. We exclude a total of 341 TNG50 pre-processed galaxies\footnote{The most massive TNG50 halo accretes a group $\mvir\sim10^{13}\, \msun$ at $z\approx0.7$, and we therefore exclude many of this group's pre-processed jellyfish.}. 

In general, we classify all TNG50 $z=0$ galaxies at all snapshots as one of the three above categories. Then for all $z=0$ systems, we further flag and exclude current backsplash galaxies, i.e. galaxies that have spent $N_{\rm snaps} = 5$ consecutive snapshots in a host of mass $\mvirhost > M_{\rm Lowlim} = 10^{11}\, \msun$ before eventually being a central galaxy at $z=0$. This definition is nearly equivalent to that used by \cite{Zinger2020}, except that they use $N_{\rm snaps} = 3$ and $M_{\rm Lowlim} = 0$ (i.e., no criterion for host mass). Additionally we note that, especially during massive mergers, \subfind may confuse which galaxy is actually the central and which is the satellite. Consequently some galaxies (such as central galaxy of the most massive cluster in TNG50) may be classified as a backsplash galaxy due to this ``swapping problem", so we recommend using caution when physically interpreting these backsplash galaxies. 

Further, we check whether each $z=0$ satellite subhalo was pre-processed or previously a backsplash galaxy. The pre-processing definition is above. A galaxy was previously a backsplash galaxy if it spent at least $N_{\rm snaps} = 5$ in any host of mass $\mvirhost > M_{\rm LowLim} = 10^{11}\, \msun$, then spent at least $N_{\rm snaps}$ consecutive snapshots as a central, before eventually being a satellite at $z=0$. A given $z=0$ satellite may be a previous backsplash but not pre-processed if the previous host is also the $z=0$ host; pre-processed but not previously a backsplash if the galaxy falls into a pre-processing host and this pre-processing host falls into the $z=0$ host; both a previous backsplash and pre-processed; or neither. Previously \citet{Donnari2021} combined these two flags -- pre-processed and previous backsplash -- as one general ``pre-processing" flag, while in this analysis we include previous backsplash galaxies. Additionally, \citet{Donnari2021} use $N_{\rm snaps} = 3$ and $M_{\rm LowLim} = 10^{12}\, \msun$, and the catalogs utilize \sublink\ rather than \sublinkgal \citep{Rodriguez-Gomez2015}.

\section{Comparisons of the Onset of RPS} \label{app:tau}

\begin{figure*}
    \includegraphics[width=\textwidth]{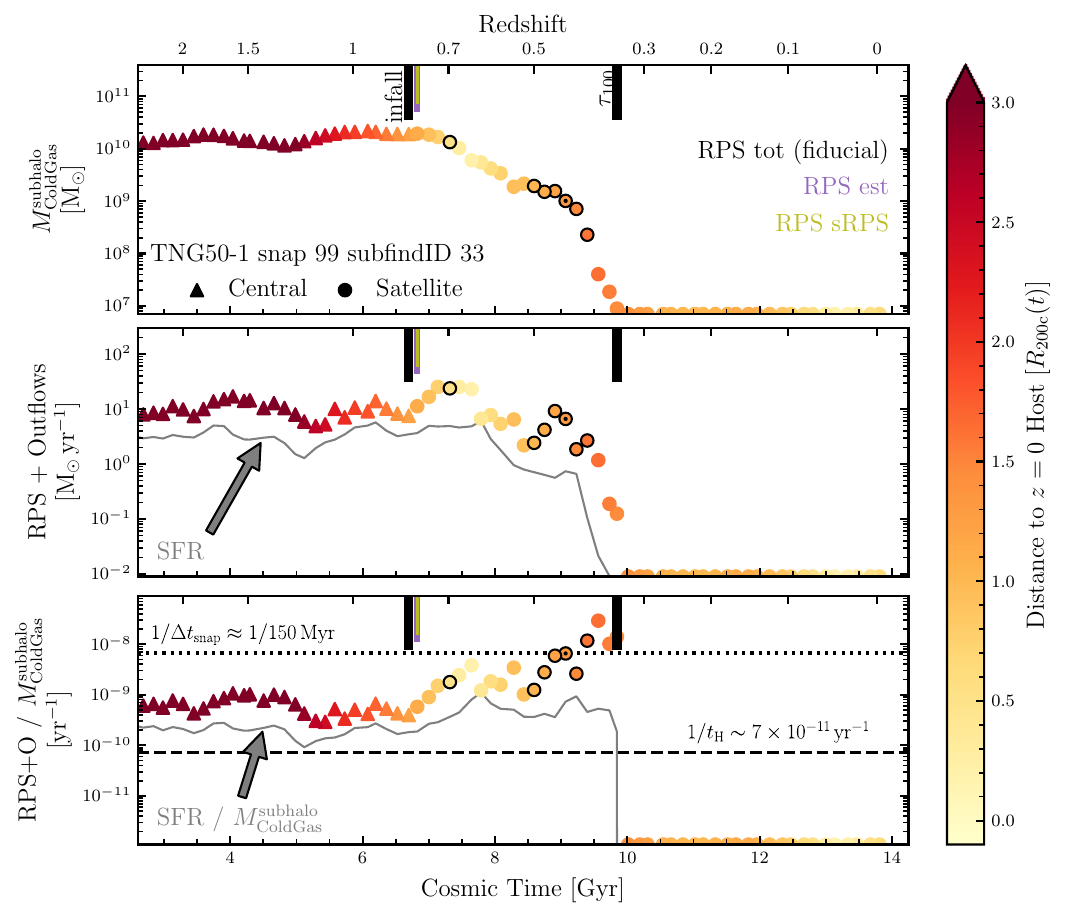}
    \caption{
    {\bf Time evolution of the cold gas content and associated tracers of a TNG jellyfish galaxy that is stripped of all cold gas between first and second pericentric passage.}
    The marker style denotes the FoF membership and the color the distance to the $z=0$ host in units of [$\rvirhost(t)$]. The inspected snapshots are outlined, and the jellyfish classified have a black dot. We plot snapshots where the y-axis quantity is below our resolution limit at the lower y-limit (along the bottom x-axis). The thick black ticks denote the fiducial start (infall) and end (when $\mcgassub = 0$) times of ram-pressure stripping (RPS), while the purple (``RPS est") and olive (``RPS sRPS") ticks denote two alternative methods of measuring the start of RPS. See the text for additional details regarding the definitions of RPS est and RPS sRPS. For this galaxy, the three definitions yield similar results for the onset of RPS. Top panel: the total gravitationally-bound cold gas mass $\mcgassub$. Middle panel: RPS+outflows and instantaneous SFR as the thin gray curve. Bottom Panel: sRPS = RPS+outflows / $\mcgassub$, in addition to SFR / $\mcgassub$ as the thin gray curve; the dashed line denotes $1 / t_H$, where $t_H$ is the Hubble Time; the dotted line denotes the approximate inverse time between snapshots $1 / \Delta t_{\rm snap} \approx 1 / 150\, {\rm Myr} \approx 7\times10^{9}\, {\rm yr}^{-1}$.
    }
\label{fig:subfindID33_MCGas_tracer_evolution}
\end{figure*}

\begin{figure*}
    \includegraphics[width=\textwidth]{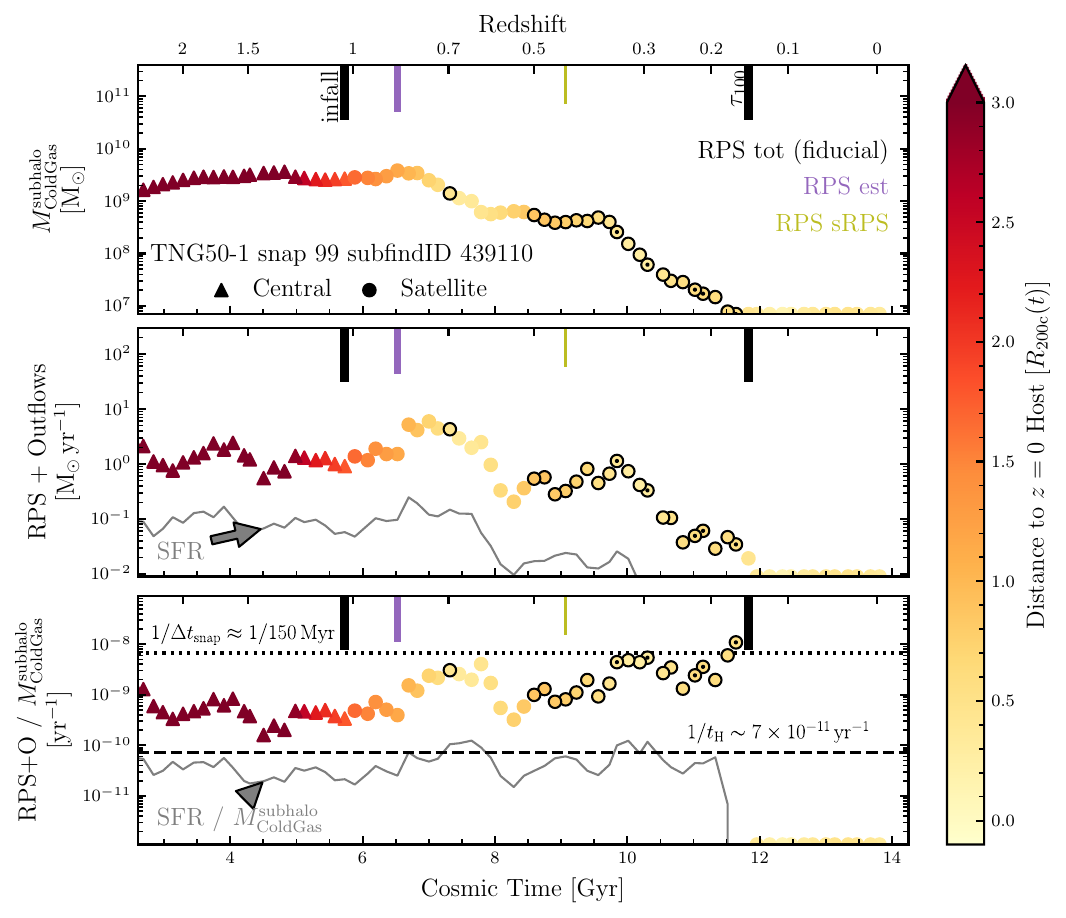}
    \caption{
    {\bf Time evolution of the cold gas content and associated tracers of a TNG50 jellyfish galaxy that is stripped of all cold gas only after its second pericentric passage.}
    Similar to Fig.~\ref{fig:subfindID33_MCGas_tracer_evolution}, except for this example the onset of RPS $\tau_0$ for ``RPS est" (purple) and ``RPS sRPS" (olive) are significantly later than the infall time (black, fiducial). This represents a non-common case, so that we can safely take the infall time, i.e. the first time a galaxy becomes part of the Fof of its $z=0$ host, as the onset of RPS.
    }
    \label{fig:subfindID439110_MCGas_tracer_evolution}
\end{figure*}

\begin{figure}
    \includegraphics[width=\columnwidth]{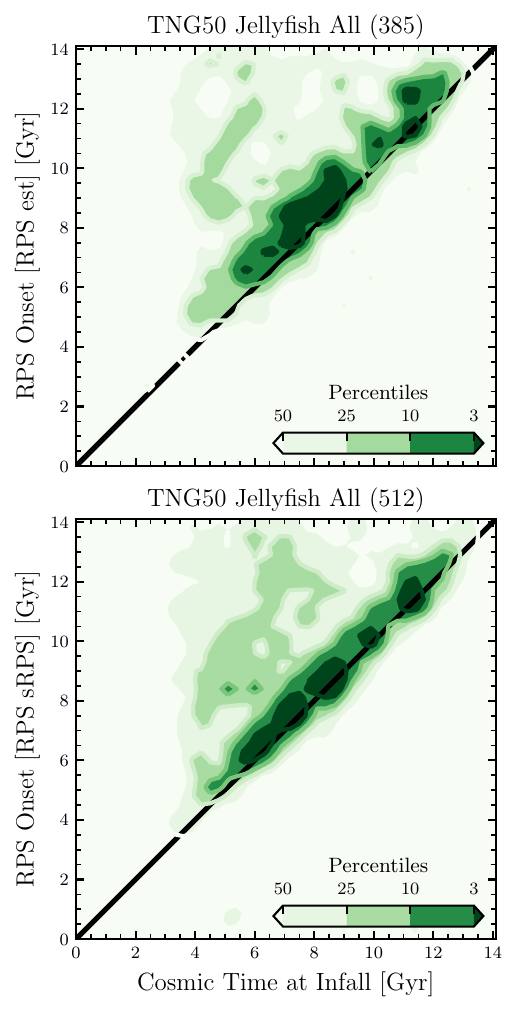}
    \caption{Alternative methods for measuring the onset of RPS compared to the fiducial choice of infall time.}
    \label{fig:tau0_contours}
\end{figure}

Throughout the paper, we define the onset $\tau_0$ of ram-pressure stripping (RPS) as the infall time, as the first time a galaxy becomes a member of its $z=0$ Friends-of-Friends (FoF) host group. The FoF algorithm decides group membership based on the relative positions of dark matter particles, and there are a-priori no constraints on the shape or total size of the halo; that is, we do not assume spherical halos where galaxies become satellite members upon crossing the virial radius $\rvir$. Consequently, there is a range of infall distances $\dsathost(\tau_0)$, ranging from $\approx 1-4\rvir$ (see Fig.~\ref{fig:fractionalRPSloss}, middle panel). Throughout this paper, we consider this FoF infall time to be onset of environmental effects for these first-infalling, not pre-processed jellyfish. Moreover, using this infall time as the onset and the time when galaxies lose all cold gas as the end of RPS allows us to measure the entire RPS time span $\tau_{\rm RPS} = \tau_{100} - \tau_{0}$, at least for the galaxies that lose all cold gas by the end of the simulation at $z=0$. 

We have checked our results using two additional definitions for measuring the onset of RPS $\tau_0$. First, we assume that the pre-infall outflows are primarily star-formation-driven outflows. Then we measure each galaxy's cold gas loading factor $\eta_{\rm ColdGas}$ as the median cold gas loss due to outflows divided by the star-formation rate (SFR)
\begin{equation} \label{eqn:eta}
    \eta_{\rm ColdGas} = \text{median}\left(\frac{\rm Outflows(<infall\, time)}{\rm SFR(<infall\, time)}\right)
\end{equation}
where ``Outflows" is the (RPS + Outflows) total cold gas mass loss from RPS + outflows directly measured using the tracer particles, before infall assumed to be entirely outflows. Then after infall, we estimate the amount of star-formation-driven outflows as the product of $\eta_{\rm ColdGas}$ and the SFR. Thus, we attempt to separate the measured RPS+Outflows into the two components:
\begin{equation}    
\begin{split}
    \text{RPS est}(t) &= (\text{RPS + outflows})(t) - \text{outflows}(t) \\
    &= (\text{RPS + outflows})(t) - \eta_{\rm ColdGas}\times \text{SFR}(t)
    \end{split}
\end{equation} 
where ``RPS est" is the estimated RPS and (RPS + outflows) is the quantity measured using the tracer particles. Then we find the peak of ``RPS est" and go backwards in time until the this estimated RPS vanishes, i.e. until the (RPS + outflows) can be fully estimated by just star-formation-driven outflows. In practice, we calculate the running median of the estimated outflows and total RPS + outflows over $N_{\rm snaps} = 7$ consecutive snapshots ($\sim 1$~Gyr), and find where the difference, the estimated RPS, peaks. Then we go backwards in time until the running median of the estimated RPS vanishes, where this time marks the onset of RPS. In Fig.~\ref{fig:subfindID33_MCGas_tracer_evolution}, the onset of RPS using this ``RPS est" is shown with a purple tick, where this estimated onset of RPS is 2~snapshots ($\approx 300$~Myr) after the infall time. In the middle panel after infall, there is an increase in the total RPS + outflows (triangles and circles) while the SFR (gray) remains approximately constant. Thus, the ``RPS est"-onset is similar to the infall time.

While this method attempts to separate the relative amounts of cold gas loss via (RPS + outflows) into RPS and outflows, there are a number of disadvantages. First, this method assumes that the cold gas mass loading factor $\eta_{\rm ColdGas}$ -- which varies with galaxy stellar mass, cold gas mass, and SFR -- is approximately constant before and after infall. Then as a number of galaxies experience a burst of star-formation between infall and first pericentric passage, the approximations of $\eta_{\rm ColdGas}$ may break down. For the example jellyfish in Fig.~\ref{fig:subfindID439110_MCGas_tracer_evolution}, there is both an increase in the RPS + outflows and in the SFR after infall, delaying the ``RPS est"-onset of RPS by 7~snapshots ($\sim1$~Gyr). In fact, for 127/512 ($\approx25$~per~cent) jellyfish galaxies, the $\eta_{\rm ColdGas}$-estimated outflows account for the entire budget of cold gas mass loss via (RPS + outflows), meaning that the estimated RPS is null. For this sample of visually-inspected galaxies to be undergoing RPS, we conclude that this method overestimates the contribution from outflows and thereby may inaccurately determine the onset of RPS. 

The increased star-formation in jellyfish galaxies during infall may be caused by the RPS-induced compression of gas, especially on the leading side \citep[e.g.,][]{Roberts2022}. Then the star-formation-driven outflows fight against the gravity of the stellar body, making the gas more susceptible to RPS \citep[e.g.,][]{Garling2022}. In this context, separating the total RPS and outflows may be futile. Both outflows and RPS work to remove the galaxy's cold gas, the fuel for star-formation, and deposit the galaxy's ISM into the halo.

In spite of this, for the remaining 385 jellyfish galaxies with non-vanishing estimated RPS contributions, the ``RPS est" onset of RPS is typically $\lesssim 3$~snapshots ($\lesssim 450$~Myr) later than the infall time (Fig.~\ref{fig:tau0_contours}, top panel). The difference between this and the fiducial onset is significantly shorter than the total RPS time span, where the end of RPS is the same for both definitions (when $M_{\rm ColdGas} = 0$). Thus, when attempting to capture the entire duration of RPS, we choose infall time over the ``RPS est" start as the fiducial onset of RPS. 

As a second alternative, we use the specific RPS and outflows (sRPS+O), namely the cold gas mass loss due to RPS and outflows (RPS+O) divided by the total amount of cold gas $\mcgassub$. Here, the units are $[{\rm time}^{-1}]$, where the inverse yields the timescale to lose all cold gas to RPS+O (at constant RPS+O). We calculate the median pre-infall sRPS+O, and find where the sRPS+O peaks. For all 259 jellyfish without cold gas at $z=0$, the maximum sRPS+O (shortest timescale) occurs at one of the last 3 snapshots that the galaxy has some cold gas. This peak value is typically $\sim 10^{-8}\, {\rm yr}^{-1}$, which is approximately the inverse time between snapshots $1 / \Delta t_{\rm snap} \approx 1/(150\, {\rm Myr})$. Then we go backwards from this peak until the running median of the sRPS+O over $N_{\rm snaps} = 7$ ($\sim 1$~Gyr) returns to the pre-infall average, and this time defines the onset of RPS. In Figs.~\ref{fig:subfindID33_MCGas_tracer_evolution},~\ref{fig:subfindID439110_MCGas_tracer_evolution}, this ``RPS sRPS" onset is marked with olive ticks. 

For galaxies that lose all cold gas approximately within the first orbit or by the first pericentric passage, such as the example in Fig.~\ref{fig:subfindID33_MCGas_tracer_evolution}, the ``RPS sRPS" onset is typically $\approx\pm 2$~snapshots ($\approx\pm 300$~Myr) of the infall time (Figure~\ref{fig:tau0_contours}, top panel). 
While the sRPS+O generally increases between infall and pericentric passage, the sRPS+O may plateau or decrease near apocenter. Sometimes this apocentric decrease may bring the sRPS+O back to the pre-infall average. In these cases, then the ``RPS sRPS" onset occurs on the second infall, such as for the example in Fig.~\ref{fig:subfindID439110_MCGas_tracer_evolution}. This leads to a number of galaxies with ``RPS sRPS" onsets significantly after infall, shown in Fig.~\ref{fig:tau0_contours} (bottom panel). This definition estimates that these galaxies actually undergo multiple periods of ram-pressure stripping. However for determining the entire duration of RPS, splitting the entire RPS process into multiple periods is not helpful. Thus, we choose the infall time as the onset of RPS.

\bsp	
\label{lastpage}
\end{document}